\documentclass{aa}
\usepackage{natbib}
\usepackage{amsmath}
\usepackage{amssymb}
\usepackage{graphics}
\usepackage[para,online,flushleft]{threeparttable}
\usepackage{tikz}
\usepackage{color, colortbl}
\usetikzlibrary{shapes,arrows}
\usetikzlibrary{shapes,snakes}

\begin{document}

\title{A near-infrared interferometric survey of debris-disk stars.\\VI. Extending the exozodiacal light survey with CHARA/JouFLU}

\author{P. D. Nu\~nez\inst{1}\thanks{paul.nunez@jpl.nasa.gov}, N. J. Scott\inst{2},  B. Mennesson\inst{3}, O. Absil\inst{4},  J.-C. Augereau\inst{5}, G. Bryden\inst{3}, T. ten Brummelaar\inst{6}, S. Ertel\inst{7}, V. Coud\'e du Foresto\inst{8}, S. T. Ridgway\inst{9}, J. Sturmann\inst{6},  L. Sturmann\inst{6}, N. J. Turner\inst{3}, N. H. Turner\inst{6}}
\institute{ NASA Postdoctoral Program Fellow, Jet Propulsion Laboratory, California Institute of Technology, 4800 Oak Grove Drive, Pasadena, CA, 91109, USA. \and NASA Ames Research Center, Moffett Field, CA 94035, USA \and Jet Propulsion Laboratory, California Institute of Technology, 4800 Oak Grove Drive, Pasadena, CA, 91109, USA. \and Space Sciences, Technologies and Astrophysics Research (STAR) Institute, Universit\'e de Li\`ege, 19c all\'ee du Six Août, 4000 Li\`ege, Belgium \and Univ. Grenoble Alpes, CNRS, IPAG, F-38000 Grenoble, France \and The CHARA Array of Georgia State University, Mount Wilson Observatory, Mount Wilson, CA 91023, USA. \and Steward Observatory, University of Arizona, 933 N Cherry Ave, Tucson, AZ 85719, USA \and Observatoire de Paris-Meudon, 5 place Jules Janssen, 92195, Meudon Cedex, France \and National Optical Astronomy Observatories, 950 North Cherry Avenue, Tucson, AZ, 85719, USA}

\titlerunning{Extending the exozodiacal light survey with CHARA/JouFLU}
\authorrunning{P. D. Nu\~nez et al.}

\abstract{
  We report the results of high-angular-resolution observations that search for exozodiacal light in a sample of main sequence stars and sub-giants. Using the ``jouvence'' of the fiber linked unit for optical recombination (JouFLU) at the center for high angular resolution astronomy (CHARA) telescope array, we have observed a total of 44 stars. Out of the 44 stars, 33 are new stars added to the initial, previously published survey of 42 stars performed at CHARA with the fiber linked unit for optical recombiation (FLUOR). Since the start of the survey extension, we have detected a K-band circumstellar excess for six new stars at the $\sim 1\%$ level or higher, four of which are known or candidate binaries, and two for which the excess could be attributed to exozodiacal dust. We have also performed follow-up observations of 11 of the stars observed in the previously published survey and found generally consistent results. We do however detect a significantly larger excess on three of these follow-up targets: Altair, $\upsilon$ And and $\kappa$ CrB. Interestingly, the last two are known exoplanet host stars. We perform a statistical analysis of the JouFLU and FLUOR samples combined, which yields an overall exozodi detection rate of $21.7^{+5.7}_{-4.1}\%$. We also find that the K-band excess in FGK-type stars correlates with the existence of an outer reservoir of cold ($\lesssim 100\,$K) dust at the $99\%$ confidence level, while the same cannot be said for A-type stars.\\
}


\maketitle

\section{Introduction}

\begin{figure}
  \begin{center}
    \scalebox{0.35}{
    \begin{tikzpicture}
      \node(image){\includegraphics[scale=2.5]{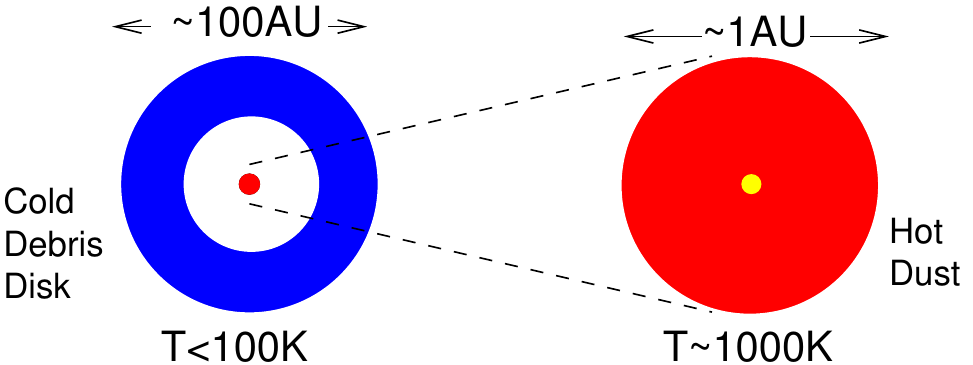}};
    \end{tikzpicture}
  }
  \end{center}
  \caption{\label{sketch} K-band excesses detected with interferometry possibly originate from hot ($\sim 1000$K) dust within an AU from the star. The left sketch shows the typical field of view attainable with a single telescope, e.g., Spitzer \citep{su_2013} and Hershel \citep{eiroa_2013}, which can resolve the outer cold-debris disk, while the magnified sketch shows the interferometric FOV, which can resolve the inner hot component of the circumstellar disk.}
\end{figure}

Observations have recently confirmed that extrasolar planets are ubiquitous, and an important objective of several future space-based missions\footnote{See for example \citet{trauger_2016}, which discusses the use of the upcoming Wide Field Infrared Survey Telescope (WFIRST) for high-contrast imaging.} is to directly image earth-like companions in the habitable zone of main-sequence stars. Several planets, lying relatively far from the habitable zone, have been directly imaged using high-contrast techniques, and a common feature has been the prior detection of a debris disk at far-infrared (FIR) wavelengths. Planets have created gaps or warps in these cold ($\sim 10\,\mathrm{K}$) debris disks (e.g., \citealt{kalas_2005}), located as far as $\sim 100\,\mathrm{AU}$ from the star, and debris disks are now considered a sign-post for the existence of planets. Debris disks lying close to the star can also hinder the detection of exoplanets, so a primary motivation for this work is to better understand the environment closer to the habitable zone of main sequence stars, and to quantify the level of exozodiacal light originating from this region (Fig. \ref{sketch}), where the debris disk is thought to have been mostly cleared out.

The small mass, and presumably low optical depth of these debris disks, makes them difficult targets for indirect detection techniques such as radial velocity (RV) and transits \citep{stark_2013}. Spectroscopic studies, although complementary, do not easily allow the disentanglement of the light from the star from that of its immediate environment. Long-baseline infrared interferometers permit resolving the inner-most region close to the star, and the use of high-precision beam-combiners has allowed interferometrists to achieve dynamic ranges of the visibility between $\sim 100$ and $\sim 1000$ at near- and mid-infrared wavelengths, for example \citet{absil_2013, mennesson_2014}. Our own solar system contains dust, thermally emitting zodiacal light at mid-infrared (MIR) wavelengths, which motivated ground-based surveys to search for exozodiacal light using long-baseline interferometers. Interestingly, the vast majority of close-in exozodiacal light detections have been in the near-infrared (NIR) \citep{ciardi, absil_2006, akeson_2009, defrere_2012, absil_2013, ertel_2014}, which are likely due to hot ($\sim 1000\,\mathrm{K}$) dust lying within an AU from the star. 

The presence of hot-circumstellar dust, as bright as a few percent relative to the star, and thousands of times brighter than our own zodiacal light, is still poorly understood. This phenomenon has been met with some skepticism in the scientific community for two main reasons: \textit{i)} NIR exozodiacal light detections still lie very close to the limit of detectability  of long-baseline interferometers, and detections are generally claimed in the few sigma range. \textit{ii)} hot and small dust particles, lying close to the star, should have a short lifetime because Poynting-Robertson drag and radiation pressure should efficiently remove dust within time-scales of a few years. Also, collision rates of larger bodies are too high for them to survive over the age of the stars around which dust is observed \citep{wyatt_2008}. Therefore, if hot dust is the origin of the NIR exozodi detections, there needs to be an efficient dust-replenishment or dust-trapping mechanism. 

In this work we will mainly discuss the recent survey results of K-band observations performed with the ``jouvence'' of the fiber linked unit for optical recombination (JouFLU) \citep{nic} at the Center for High Angular Resolution Astronomy (CHARA) Array \citep{chara}, and combine our results with the survey performed by \citet{absil_2013}. To establish confidence in the results, and to deal with item \textit{(i)}, we first developed a data analysis strategy for minimizing the bias and uncertainty of exozodiacal light observations with two-beam Michelson interferometers, which is described in detail in \citet{nunez_2016}. Using this analysis strategy we also performed a study of calibrator stars to verify that the measured circumstellar excesses are not a result of systematic errors. Here we present the full survey results, explore possible correlations with basic stellar parameters, and discuss the plausibility of different physical mechanisms that may be responsible. As mentioned above in item \textit{(ii)}, the main difficulty is to find a physical mechanism that can be responsible for the hot dust phenomenon. Some explanations involve comet in-fall or out-gassing \citep{wyatt_2008, su_2013, lebreton_2013, bonsor_2012}. Another explanation proposed by \citet{su_2013}, and further developed by \citet{rieke_2016}, is that magnetic trapping of small enough dust grains (nano-grains), close to the star's sublimation radius, is possible with typical magnetic fields of main sequence stars, and predicts that detection rates should increase with stellar rotational velocities. We search for the predicted correlations in our sample.

The outline of this paper is the following: in Sections \ref{stellar_sample} and \ref{strategy} we discuss the JouFLU stellar sample and observing strategy. In Sections \ref{vis_estimation} and \ref{csf_estimation} we discuss the interferometric visibility and circumstellar emission level estimation procedures. In Sections \ref{results} and \ref{follow_up}, we report the measured circumstellar emission levels for the JouFLU survey extension and follow-up observations of excess stars from the original \citet{absil_2013} survey. In Section \ref{statistical_analysis}, we perform a statistical analysis of the measured excess levels, and explore correlations with basic stellar parameters. 

\section{The JouFLU stellar sample} \label{stellar_sample}

Since 2013, we have extended the initial K-band survey of 42 stars performed by \citet{absil_2013}. The main selection criteria were that targets should have not departed significantly from the main sequence and that they should not have known multiplicity. In addition, we aimed to have a target list that covers many spectral types, and as many stars with known cold dust reservoirs (MIR-FIR excess) as without. However, the main restriction is that targets be bright enough (K$\lesssim 4.2$) to achieve sub-percent precision on the squared visibility. There are approximately $\sim 214$ main sequence stars with no known multiplicity and brighter than $K=4.2$ observable at CHARA.

\begin{figure}
  \begin{center}
    \includegraphics[scale=0.45]{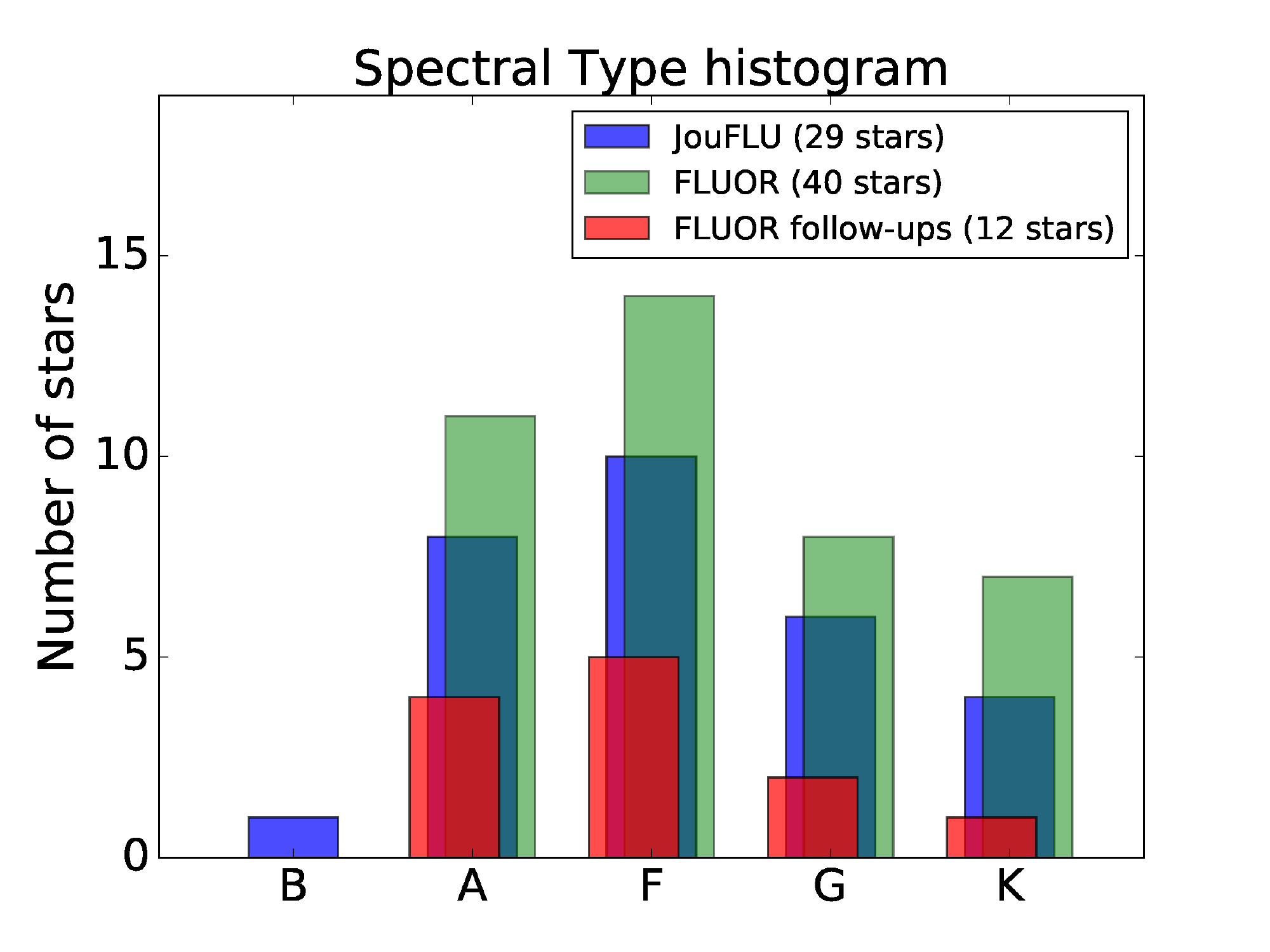}
  \end{center}
  \caption{\label{stype_histo} Distribution of spectral types for the stars in the \citet{absil_2013} survey (green),  the stars in the JouFLU survey (blue) in this work, and the FLUOR follow-up stars (red).}
\end{figure}

\begin{figure}
  \begin{center}
    \includegraphics[scale=0.45]{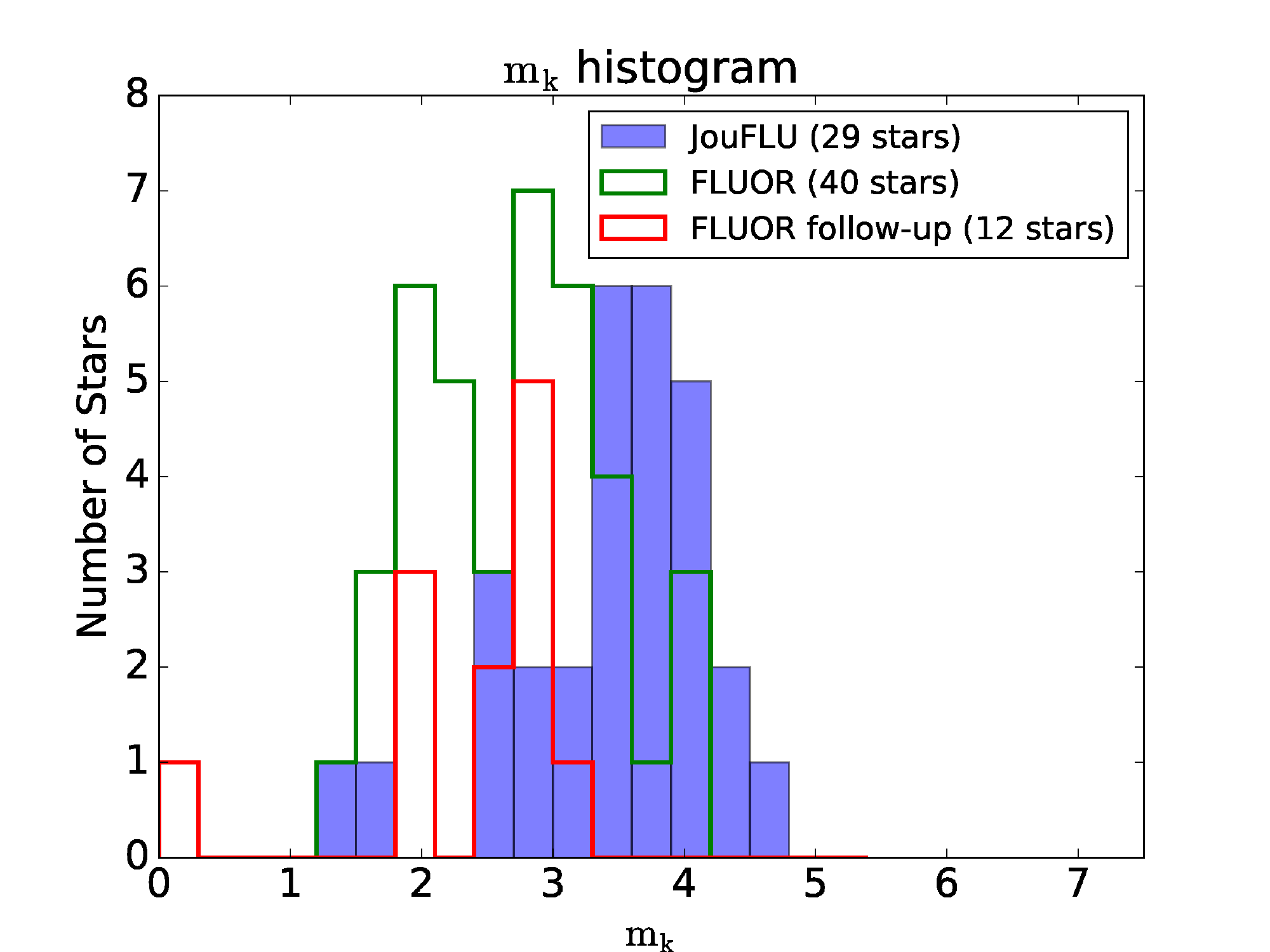}
  \end{center}
  \caption{\label{mk_histo} Histogram of K-band magnitude for the stars in the \citet{absil_2013} FLUOR survey (green), the JouFLU survey (blue, this work), and the FLUOR follow-up stars (red).}
\end{figure}

We have added 33 new stars to the survey as shown in Table \ref{survey_table}, distributed across many spectral types as shown in Fig. \ref{stype_histo}. In comparison with the initial FLUOR sample of \citet{absil_2013}, which had a median K magnitude of 2.9, the JouFLU stellar sample is somewhat fainter, as shown in Fig. \ref{mk_histo}, with a median K magnitude of 3.5. The FLUOR sample contains 19 stars with either a MIR or FIR excess, which is attributed to a cold-dust ($\lesssim 100\,$K) reservoir. The JouFLU stellar sample only contains five stars with a such a suspected cold-dust reservoir, mainly because these stars are less common and quickly exceed the current limiting magnitude attainable by JouFLU (K$\sim 4.5$). 

The JouFLU sample contains eight targets\footnote{Common targets with the VLTI's survey with the Precision Integrated Optics Near Infrared ExpeRiment (PIONIER) are: HD23249, HD28355, HD33111, HD164259, HD165777, HD182572, HD210418, HD215648.} in common with another H-band exozodi survey conducted in the southern hemisphere with the Very Large Telescope Interferometer (VLTI) \citep{ertel_2014}.  We also note that the JouFLU sample also has eight common targets\footnote{Common targets with the LBTI survey are: HD33111, HD26965, HD19373, HD182572, HD182640, HD185144, HD210418, HD219134.} with the Large Binocular Telescope Interferometer survey \citep{weinberger_2015}, which will enable a comparison with high resolution and high-contrast mid-infrared observations.

\begin{table*}
\caption{\label{survey_table}New stars observed between 2013 and 2015 for the JouFLU exozodi survey extension. Shown here are the HD number, common name, K-band magnitude, Age, baseline used, and date of observation.}
\begin{tabular}{ccccccc}
\hline \hline
HD & Name & Type & K & Age (Gyr) & Baseline & Date \\ \hline 
5015 & HR 244 & F9V & 3.54 & 5.3 & S1-S2 & 08/15/2015, 9/23/2015\\ \hline 

5448 & 37 And & A5V & 3.5 & 0.6 & S1-S2 & 8/13/2015, 8/14/2015\\ \hline 

6961 & $\theta$ Cas & A7V & 3.97 & ? & S1-S2 & 08/15/2015, 9/23/2015\\ \hline 

10780 & V987 Cas & K0V & 3.84 & 2.3 & S1-S2 & 5/25/2015\\ \hline 

14055 &  $\gamma$ Tri & A1Vnn & 3.958 & 0.3 & E1-E2 & 10/11/2015, 10/14/2013\\ \hline 

15335 & 13 Tri & G0V & 4.48 & 6.45 & E1-E2 & 10/16, 10/19/2013\\ \hline 

19373 & $\iota$ Per & F9.5V & 2.685 & 4.1 & S1-S2 & 10/14/2014\\ \hline 

20630 &  $\kappa$1 Cet & G5Vv & 3.34 & 0.35 & E1-E2 & 10/17/2013, 10/18/2013\\ \hline 

23249 & $\delta$ Eri & K1III-IV & 1.43 & 5.9 & E1-E2 & 10/11/2013, 10/14/2013\\ \hline 

26965 & 40 Eri & K0.5V & 2.498 & 5.6 & S1-S2 & 10/14/2014\\ \hline 

28355 & 79 Tau & A7V & 4.55 & 0.67 & E1-E2 & 10/12/2013, 10/18/2013\\ \hline 

33111 & $\beta$ Eri & A3IV & 2.38 & 0.39 & S1-S2 & 9/23/2015, 9/24/2015, 9/26/2015\\ \hline 

34411 & $\lambda$ Aur & G1.5IV-V & 3.27 & 5.3 & E1-E2 & 10/16/2013, 10/17/2013\\ \hline 

50635 & 38 Gem & F0Vp & 3.89 & 4.98 & E1-E2 & 10/15/2013, 10/16/2013\\ \hline 

87901 & $\alpha$ Leo & B8IVn & 1.599 & 1.24 & S1-S2 & 11/11/2014\\ \hline 

162003A & HR 6636 & F5IV-V & 3.43 & 3.8 & E1-E2 & 07/14/2014\\ \hline 

164259 & $\zeta$ Ser & F2IV & 3.67 & 2.4 & E1-E2 & 6/15/2015 6/16/2015\\ \hline 

165777 & 72 Oph & A4IVs & 3.42 & 0.55 & E1-E2 & 6/19/2015, 6/20/2015\\ \hline 

168151 & LTT 15404 & F5V & 3.85 & 2.5 & E1-E2 & 7/14/2014\\ \hline 

182572 & 31 Aql & G8IV... & 3.53 & 4.5 & E1-E2 & 7/23/2015, 6/15/2015, 6/20/15\\ \hline 

182640 & $\delta$ Aql & F0IV & 2.52 & ? & S1-S2 & 10/14/2014\\ \hline 

184006 & $\iota$ Cyg & A5V & 3.598 & 0.45 & E1-E2 & 10/12/2013\\ \hline 

187691 & LTT 15798 & F8V & 3.902 & 3.3 & E1-E2 & 06/20/2015\\ \hline 

190360 & LHS 3510 & G7IV-V & 4.05 & 12.1 & E1-E2 & 10/16/2013, 10/19/13\\ \hline 

192640 &  V1644 Cyg & A2V & 4.47 & 1.0 & E1-E2 & 10/17/2013, 10/18/13\\ \hline 

202444 &  $\tau$ Cyg & F0IV & 2.69 & ? & S1-S2 & 9/26/2015\\ \hline 

210418 & $\theta$ Peg & A1Va & 3.33 & 0.8 & E1-E2 & 10/12/2013, 10/14/13\\ \hline 

213558 & $\alpha$ Lac & A1V & 3.75 & 0.4 & E1-E2 & 10/11/2013,  10/12/13\\ \hline 

215648 & LHS 3851 & F7V & 2.92 & 6.7 & S1-S2 & 8/13/2015\\ \hline 

217014 & 51 Peg & G2.5IVa & 3.99 & 7.1 & E1-E2 & 10/16/2013, 10/19/13\\ \hline 

219080 & 7 And & F0V & 3.77 & 2.6 & S1-S2 & 9/24/2015\\ \hline 

219134 & HR 8832 & K3V & 3.125 & 12.5 & S1-S2 & 10/14/2014\\ \hline 

222368 & $\iota$ Psc & F7V & 2.825 & 5.2 & E1-E2 & 10/17/2013, 10/18/2013\\ \hline 

\end{tabular}
\end{table*}

In addition to the JouFLU survey, we performed follow-ups on 11 stars of the \citet{absil_2013} FLUOR sample in order to confirm some previous detections (see Table \ref{follow_up_table}), and search for possibly time variability of the exozodiacal light level.

\begin{table*}
\caption{\label{follow_up_table}Stars observed between 2013 and 2015 for the exozodi follow-up program. Shown here are the HD number, common name, K-band magnitude, Age, baseline used, and date of observation. HD185144 is a common target with the LBTI survey.}
\begin{tabular}{ccccccc}
\hline\hline
HD & Name & Type & K & Age (Gyr) & Baseline & Date \\ \hline 
9826 & $\upsilon$ And & F9V & 2.84 & 4.0 & S1-S2 & 9/16/2016, 9/17/2016, \\ \hline 


40136 &  $\eta$ Lep & F2V & 2.911 & 2.3 & S1-S2 & 10/13/2014, 11/10/2014\\ \hline 

102647 & $\beta$ Leo & A3Va & 1.915 & 0.1 & S1-S2 & 5/1/2015, 5/27/2015, 5/28/2015\\ \hline 

131156 & $\xi$ boo & G7V & 1.97 & 0.3 & S1-S2 & 02/15/2015\\ \hline 

142091 & $\kappa$ CrB & K1IV & 2.49 & 2.5 & S1-S2, E1-E2 & \tiny{5/1/2015, 6/18/2015, 6/19/2015, 6/20/2015}\\ \hline 

142860 & $\gamma$ Ser & F6IV & 2.62 & 4.6 & E1-E2 & 5/15/2013, 5/25/2014, 6/16/2015\\ \hline 

173667 &  110 Her & F6V & 3.05 & 3.4 & S1-S2 & 06/16/2015\\ \hline 

177724 & $\zeta$ Aql & A0Vn & 2.9 & 0.8 & E1-E2 & 06/18/2015\\ \hline 

185144 & $\sigma$ Dra & G9V & 2.807 & 3.2 & S1-S2 & 08/15/2015\\ \hline 

187642 & $\alpha$ Aql & A7V & 0.219 & 1.3 & S1-S2 & 7/23/2014, 6/17/2015\\ \hline 

203280 & $\alpha$ cep & A7IV & 1.96 & 0.8 & E1-E2 & 2015/8/13, 8/15/2015\\ \hline 
\end{tabular}
\end{table*}

\section{Observing strategy} \label{strategy}

We used the JouFLU beam combiner at the CHARA array to perform observations. JouFLU, an upgraded version of FLUOR \citep{fluor1, fluor2}, is a two-telescope beam combiner which uses single-mode fibers and photometric channels to enable precise visibility measurements at the $\sim 1\%$ level on bright (K$\lesssim 3$) stars \citep{nic, fluor1, fluor2}. Interference fringes are scanned at  high frequency (100 Hz) in order to minimize the effects due to atmospheric piston, and we nominally collect 150-200 fringe scans for each target or calibrator star. To minimize calibration biases, we generally use three different calibrator stars (cals) for each science target, hereafter ``obj''. Cal$_1$ and Cal$_2$ have respectively a lower and higher right-ascension than the target, and Cal$_3$ is located as close as possible to the target. An observing block is generally of the form (cal$_1$, obj, cal$_{2}$, obj, cal$_3$, obj, cal$_1$). We generally obtain approximately six calibrated points (two observing blocks) for each science target. 

We have mainly used the smallest baselines of the CHARA array S1-S2 ($33\,\mathrm{m}$) and E1-E2 ($60\,\mathrm{m}$), for which stars remain mostly unresolved, and with intrinsic visibilities that hence have a very weak dependence on the assumed stellar model diameter. For example, the largest and brightest (K=2) calibrator used in the survey has a uniform angular diameter estimated at $1.71 \pm 0.024\,\mathrm{mas}$ \citep{merand_2005}. This corresponds to an interferometric visibility of $0.979\pm 0.0006$ at the $33\,\mathrm{m}$ baseline. Even assuming a more conservative diameter uncertainty of 5\% on this worst case calibrator, the resulting visibility uncertainty is 0.7\%. In fact, V-K surface brightness angular diameters are sufficient to achieve the sub-percent precision level required for the survey. In \citet{nunez_2016} we also show that the difference in brightness between the target and the calibrators does not bias the visibility measurements to levels higher than $\sim 0.1\%$, particularly when we use the visibility estimation and the calibration methods described in the above reference, which is used for all the data reported here. \\

\section{Visibility estimation procedure} \label{vis_estimation}

We start by obtaining raw (uncalibrated) visibilities: from the ensemble of fringe scans obtained for the science target and calibrators, we computed the median visibility and bootstrapped uncertainty. We then calibrated the raw visibilities by estimating the transfer function at the time of the science target observation using the ``hybrid interpolation'' as described in \citet{nunez_2016}. The final uncertainty in the calibrated visibility takes into account the statistical error from the median bootstrapping, and the systematic uncertainty from changes in the transfer function. This procedure is discussed in detail in \citet{nunez_2016}.

\section{Exozodiacal light-level estimation procedure} \label{csf_estimation}

\begin{figure}
  \begin{center}
    \includegraphics[scale=0.35, angle=-90]{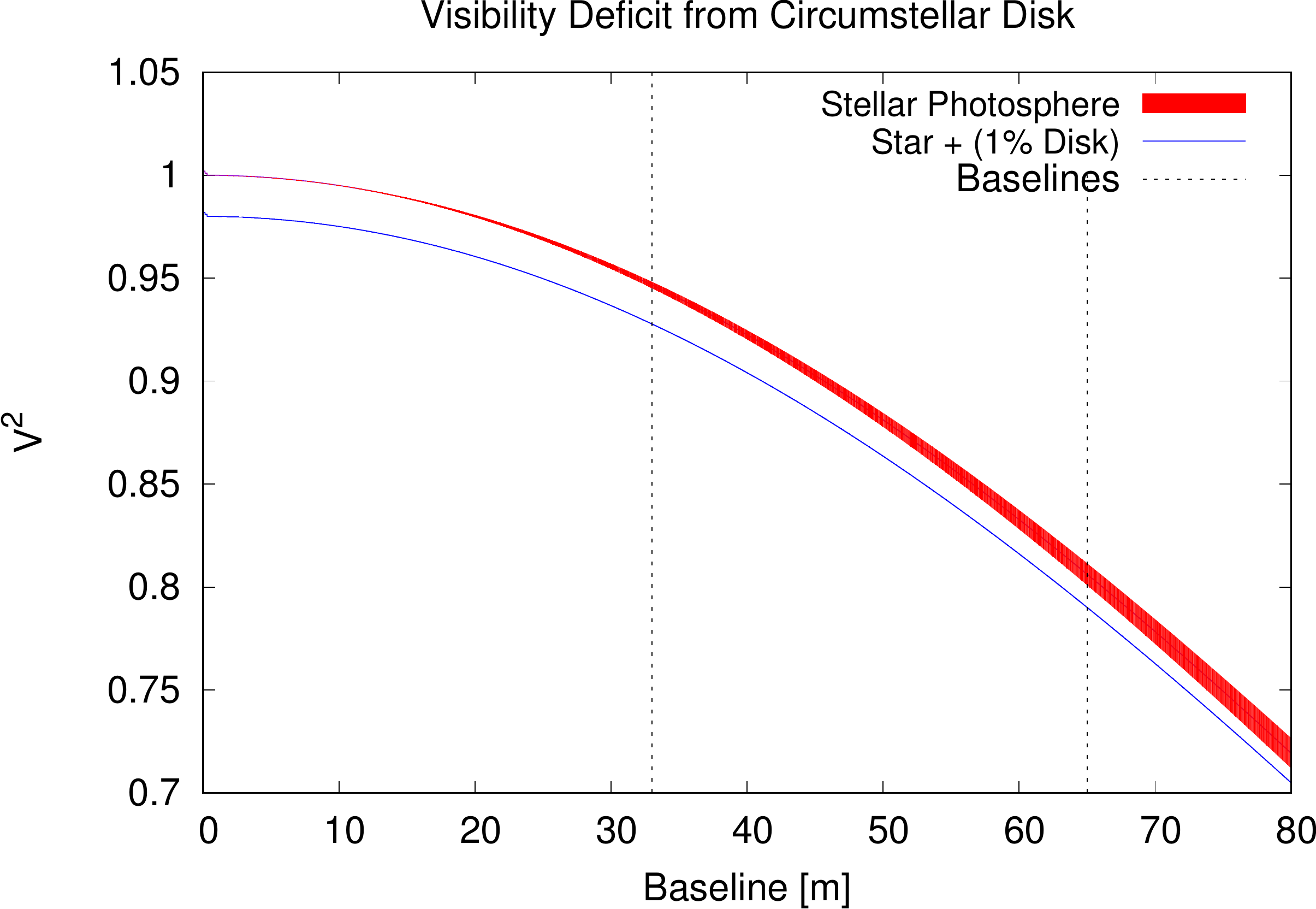}
  \end{center}
  \caption{\label{detection_method} Filled red curve represents the predicted squared visibility for a uniform-disk type star, with a diameter of $(1\pm0.05)\,\mathrm{mas}$. The blue line represents the squared visibility of the star with a circumstellar disk of $1\%$ brightness relative to the star, that uniformily fills the entire field of view. We note that at short baselines, the uncertainty of the stellar model, i.e., the width of the red curve, is much smaller than the visibility deficit that results from fully resolving the faint circumstellar disk. The squared visibility is equal to one at zero baseline for both models. Also shown are the 33$\,$m and 65$\,$m baselines used for this study.}
\end{figure}

The main idea behind the interferometric detection of exozodiacal light is that the star remains mostly unresolved at short baselines, while exozodiacal light emission spreads over a much bigger region and is partially or fully resolved by the interferometer. Faint off-axis emission is therefore detected as a small decrease of the fringe visibility (contrast) at short baselines as shown in Fig. \ref{detection_method}. If exozodiacal light uniformly fills the entire FOV, then the circumstellar emission level $f_{\mathrm{cse}}$ relative to the stellar emission level, is approximately related to the squared fringe visibility as \citep{di_folco_2007}

\begin{equation}
  V^2 = V_{\star}^2(1-2f_{\mathrm{cse}}), \label{fcse}
\end{equation}

where $V_{\star}$ is the expected visibility of the stellar photosphere. This is clearly a simple model, but it allows estimating the circumstellar excess within the FOV when no information  is available about the spatial distribution of the excess \citep{absil_2006}. To estimate the expected photosphere visibility, we used interferometric stellar radii and linear limb-darkenned coefficients when available. When stellar radii were not directly available, we used surface brightness relations to estimate the limb darkened angular diameter as described in \citet{groenewegen_2004}. Next, we performed a fit of $f_{\rm{cse}}$ in Eq. \ref{fcse}, which is the only free parameter of the model. To estimate the uncertainty $\sigma_{\rm{cse}}$ on the circumstellar emission level relative to the star, we identified the values of $f_{\rm{cse}}$ that increase the best fit $\chi^2$ by 1. An excess detection is claimed when $f_{cse}/\sigma_{cse}\geq 3$.

A major strength of this detection strategy is that it is largely insensitive to the uncertainty of the stellar model when observations are performed at short baselines as shown in Fig \ref{detection_method}, and discussed in Section 4 of \citet{nunez_2016}. Also, while it is possible that calibrator stars display a circumstellar excess, this would tend to underestimate any excess found around the science target, and tend to decrease the number of detections. Since we used several (approximately three) calibrator stars for each science target, we further verified that the measured excesses were not an effect of systematic errors by checking that the calibrators can be used to calibrate each other (see Section \ref{cal_analysis}).

\section{Results of circumstellar excess levels} \label{results}

For each target we tested two main models: a stellar photosphere model without circumstellar disk, and a stellar model with a circumstellar disk as described in Section \ref{csf_estimation}. In Table \ref{csf_survey_table} we present the best-fit circumstellar emission levels for the JouFLU exozodi survey, and the reduced $\tilde{\chi^2}$ for both models.

\begin{table*}
\caption{\label{csf_survey_table}Circumstellar emission level for the exozodi survey. For each target, the table shows the identifiers, K-band magnitude, interferometric baselines used (S1-S2$\sim 33\,\mathrm{m}$, E1-E2$\sim60\,\mathrm{m}$), circumstellar emission level, reduced $\tilde{\chi}_{\mathrm{cse}}^2$ for the star + pole-on-disk model (Eq. \ref{fcse}), and reduced $\tilde{\chi}_{\star}^2$ for the photosphere model. The highlighted boxes correspond to statistically significant excesses: red boxes indicate that there is no evidence of binarity, while yellow boxes indicate that there is separate evidence of binarity.}
\begin{tabular}{ccccclcc}
\hline\hline
HD & Name & Type & K & Baseline & $f_{cse}$(\%) & $\tilde{\chi^2}_{\mathrm{cse}}$  &  $\tilde{\chi^2}_{\star}$\\ \hline 
5015 & HR 244 & F9V & 3.54 & S1-S2 & $0.61\pm0.83$ & 1.79 & 1.68\\ \hline 

5448 &  37 And & A5V & 3.5 & S1-S2 & \cellcolor{yellow!25} $2.64\pm0.53$ & 0.716 & 5.24\\ \hline 

6961 &  $\theta$ Cas & A7V & 3.97 & S1-S2 & $0.64\pm0.67$ & 3.4 & 2.91\\ \hline 

10780 & V987 Cas & K0V & 3.84 & S1-S2 & $2.34\pm0.89$ & 3.42 & 3.47\\ \hline 

14055 & $\gamma$ Tri & A1Vnn & 3.96 & E1-E2 & $-1.39\pm1.08$ & 0.891 & 1.67\\ \hline 

15335 &  13 Tri & G0V & 4.48 & E1-E2 & $0.60\pm1.41$ & 0.651 & 0.51\\ \hline 

19373 & $\iota$ Per & F9.5V & 2.68 & S1-S2 & $-0.36\pm0.28$ & 2.267 & 2.12\\ \hline 

20630 & $\kappa$1 Cet & G5Vv & 3.34 & E1-E2 & $1.39\pm1.03$ & 0.83 & 1.08\\ \hline 

23249 &  $\delta$ Eri & K1III-IV & 1.43 & E1-E2 & $1.40\pm0.77$ & 0.092 & 1.75\\ \hline 

26965 &  40 Eri & K0.5V & 2.498 & S1-S2 & $0.01\pm0.76$ & 1.94 & 1.39\\ \hline 

28355 &  79 Tau & A7V & 4.55 & E1-E2 & $-0.86\pm1.23$ & 4.157 & 3.16\\ \hline 

33111 &  $\beta$ Eri & A3IV & 2.38 & S1-S2 & \cellcolor{yellow!25} $3.21\pm0.37$ & 6.35 & 17.67\\ \hline 

34411 & $\lambda$ Aur & G1.5IV-V & 3.27 & E1-E2 & $0.25\pm0.56$ & 0.525 & 0.48\\ \hline 

50635 &  38 Gem & F0Vp & 3.89 & E1-E2 & $2.0\pm1.0$ & 3.93 & 3.55\\ \hline 

87901 & $\alpha$ Leo & B8IVn & 1.60 & S1-S2 & $-0.12\pm0.79$ & 0.539 & 0.37\\ \hline 

162003A & HR 6636 & F5IV-V & 3.43 & E1-E2 & \cellcolor{yellow!25} $7.44\pm0.52$ & 1.431 & 29.02\\ \hline 

164259 & $\zeta$ Ser & F2IV & 3.67 & E1-E2 & $0.76\pm0.88$ & 2.748 & 2.48\\ \hline 

165777 & 72 Oph & A4IVs & 3.42 & E1-E2 & $3.27\pm1.46$ & 1.46 & 1.68\\ \hline 

168151 & LTT 15404 & F5V & 3.85 & E1-E2 & $1.84\pm0.76$ & 3.30 & 3.89\\ \hline 

182572 &  31 Aql & G8IV & 3.53 & E1-E2 & $0.03\pm0.6$ & 2.34 & 2.05\\ \hline 

182640 & $\delta$ Aql & F0IV & 2.52 & S1-S2 & \cellcolor{yellow!25} $7.2\pm0.41$ & 1.91 & 51.24\\ \hline 

184006 & $\iota$ Cyg & A5V & 3.598 & E1-E2 & $-0.54\pm0.875$ & 1.581 & 1.32\\ \hline 

187691 & LTT 15798 & F8V & 3.902 & E1-E2 & $1.17\pm1.87$ & 0.48 & 0.84\\ \hline 

190360 & LHS 3510 & G7IV-V & 4.05 & E1-E2 & $0.05\pm0.52$ & 1.55 & 1.32\\ \hline 

192640 & V1644 Cyg & A2V & 4.47 & E1-E2 & $2.16\pm1.35$ & 1.667 & 1.95\\ \hline 

202444 & $\tau$ Cyg & F0IV & 2.69 & S1-S2 & $2.83\pm1.35$ & 3.44 & 3.58\\ \hline 

210418 & $\theta$ Peg & A1Va & 3.33 & E1-E2 & \cellcolor{red!25} $1.69\pm0.54$ & 1.67 & 2.80\\ \hline 

213558 & $\alpha$ Lac & A1V & 3.75 & E1-E2 & $-1.17\pm0.86$ & 0.43 & 0.71\\ \hline 

215648 & LHS 3851 & F7V & 2.92 & S1-S2 & $0.51\pm0.42$ & 1.08 & 1.17\\ \hline 

217014 & 51 Peg & G2.5IVa & 3.99 & E1-E2 & $-0.03\pm0.76$ & 0.62 & 0.51\\ \hline 

219080 & 7 And & F0V & 3.77 & S1-S2 & $-0.2\pm0.54$ & 0.98 & 0.87\\ \hline 

219134 & HR 8832 & K3V & 3.12 & S1-S2 & $0.58\pm0.54$ & 1.61 & 1.55\\ \hline 

222368 & $\iota$ Psc & F7V & 2.82 & E1-E2 & \cellcolor{red!25} $1.58\pm0.36$ & 1.41 & 4.82\\ \hline 

\end{tabular}
\end{table*}

The survey results for the new JouFLU targets, summarized in Table \ref{csf_survey_table}, show six new circumstellar excesses, including two possibly associated with circumstellar dust around HD210418 and HD222368. The other four cases (HD162003, HD182640, HD33111, and HD5448) are likely binaries, and in the cases of HD182640 and HD5448, we checked for compatibility of the data with binary orbit solutions obtained by other authors. We highlight the first direct detection of a previously unseen companion for HD162003A as discussed in Section \ref{binary_sec}. Below we present reduced data for some representative stars, which fall into three main categories: non-excess detections, excess detections attributed to binarity, and excess detections attributed to dust.

\subsection{Examples of survey stars with no excess detected: HD34411, HD190360}

\begin{figure*}
  \begin{center}
    \begin{eqnarray}
    \includegraphics[scale=0.70]{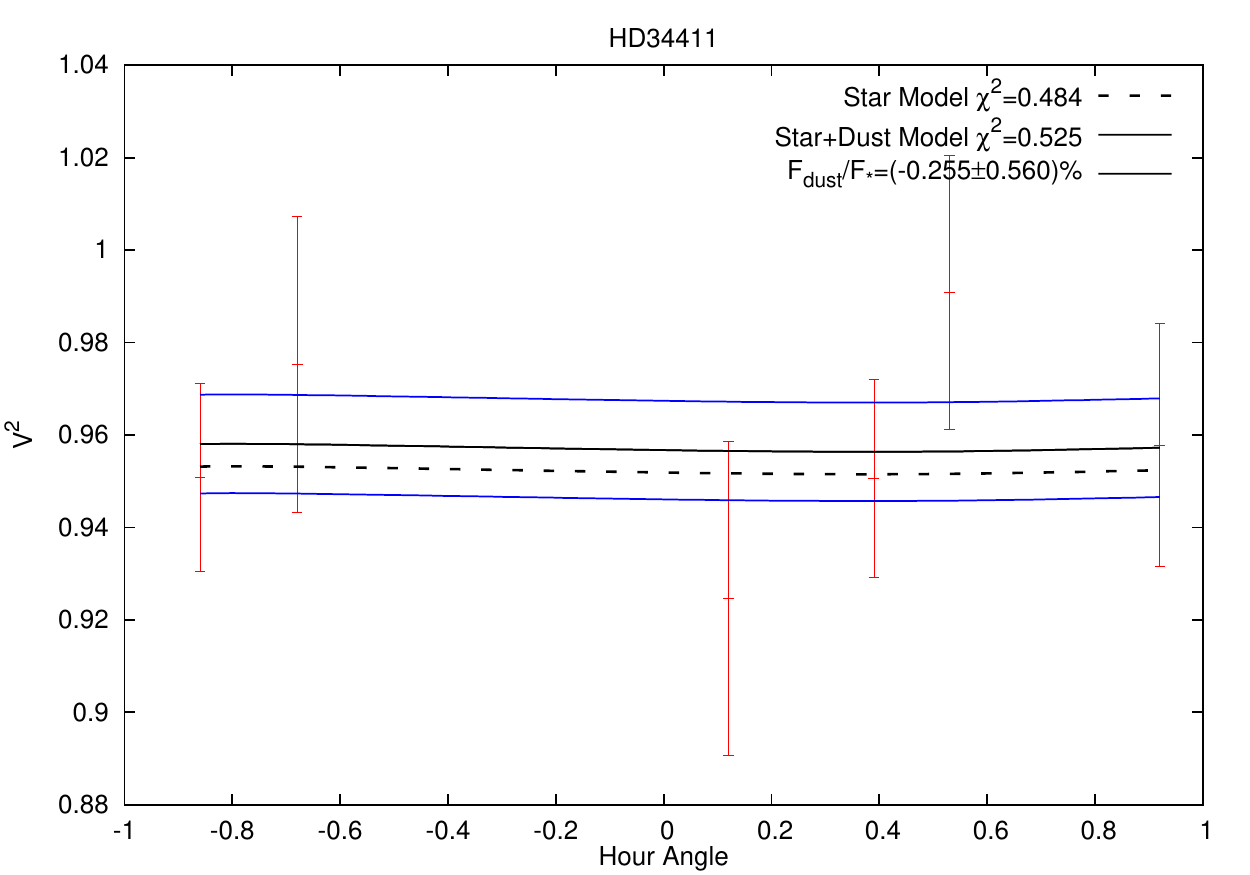} & \includegraphics[scale=0.70]{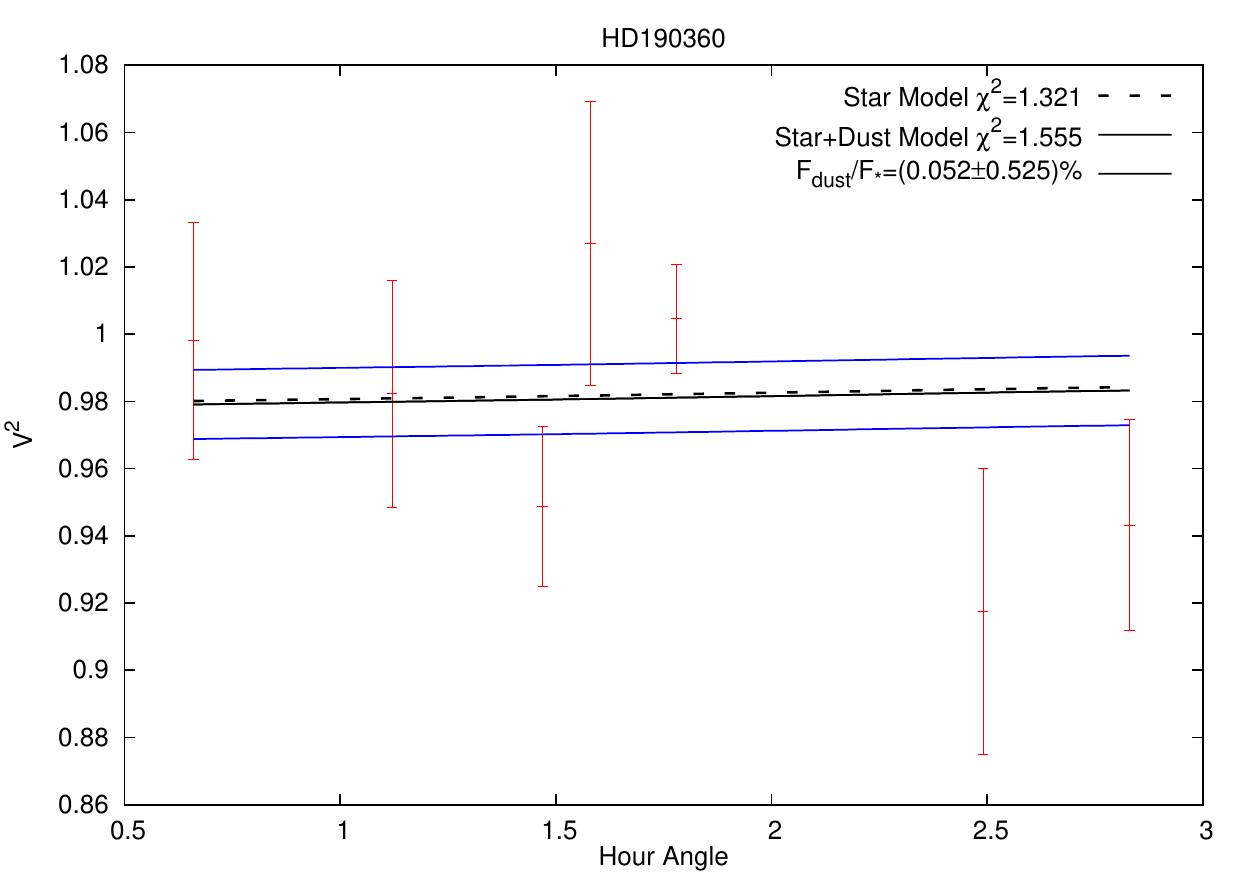} \nonumber
    \end{eqnarray}
  \end{center}
  \caption{\label{non_detections} Squared visibility as a function of hour angle for the E1-E2 ($65\,\mathrm{m}$) baseline. The dashed curve represents the stellar photosphere model, with a model uncertainty of $<0.1\%$ (not shown). The solid black curve represents the best fit squared visibility model using excess emission that uniformly fills the field of view. Both objects are consistent with a non-excess and compatible with the stellar photosphere model as indicated by the low reduced $\tilde{\chi}^2$ values of the star only model.}
\end{figure*}

\textbf{HD34411} (lam Aur) is a solar-type star with a known far-infrared ($70\mu\mathrm{m}$) excess detected with the Spitzer space telescope \citep{trilling_2008}, which implies the existence of a cold debris disk. The 2013 K-band observations with the $65\,\mathrm{m}$ baseline are consistent with a non-excess as shown in Fig. \ref{non_detections}, and are compatible with the stellar photosphere model that uses the limb darkened photospheric model derived by \citet{taby_2012}.\\

\textbf{HD190360} is a solar-type exoplanet host star \citep{naef_2003}, with two detected Jovian planets, the farthest one being $\sim 3.9\,\mathrm{AU}$ ($\sim 4.2\,\mathrm{mas}$) from the star. The K-band interferometric data, obtained with the $65\,\mathrm{m}$ baseline, are compatible with the stellar photosphere parameters derived by \citet{ligi_2016}, and therefore consistent with a non-excess as shown in Fig. \ref{non_detections}.

\subsection{Survey excess stars with no evidence for binarity: HD210418, HD222368} \label{survey_detections}

\begin{figure*}
  \begin{center}
    \begin{eqnarray}
        \includegraphics[scale=0.7]{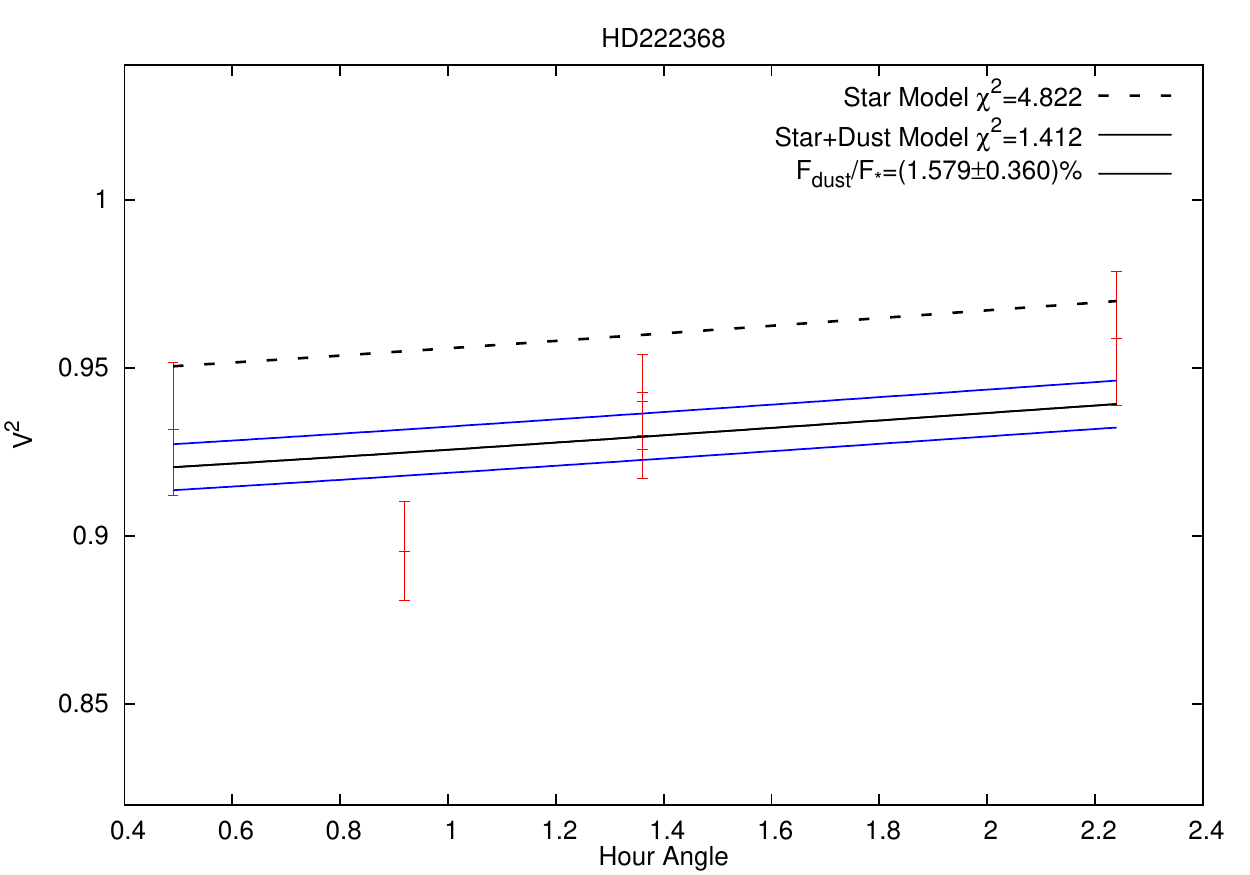} & \includegraphics[scale=0.7]{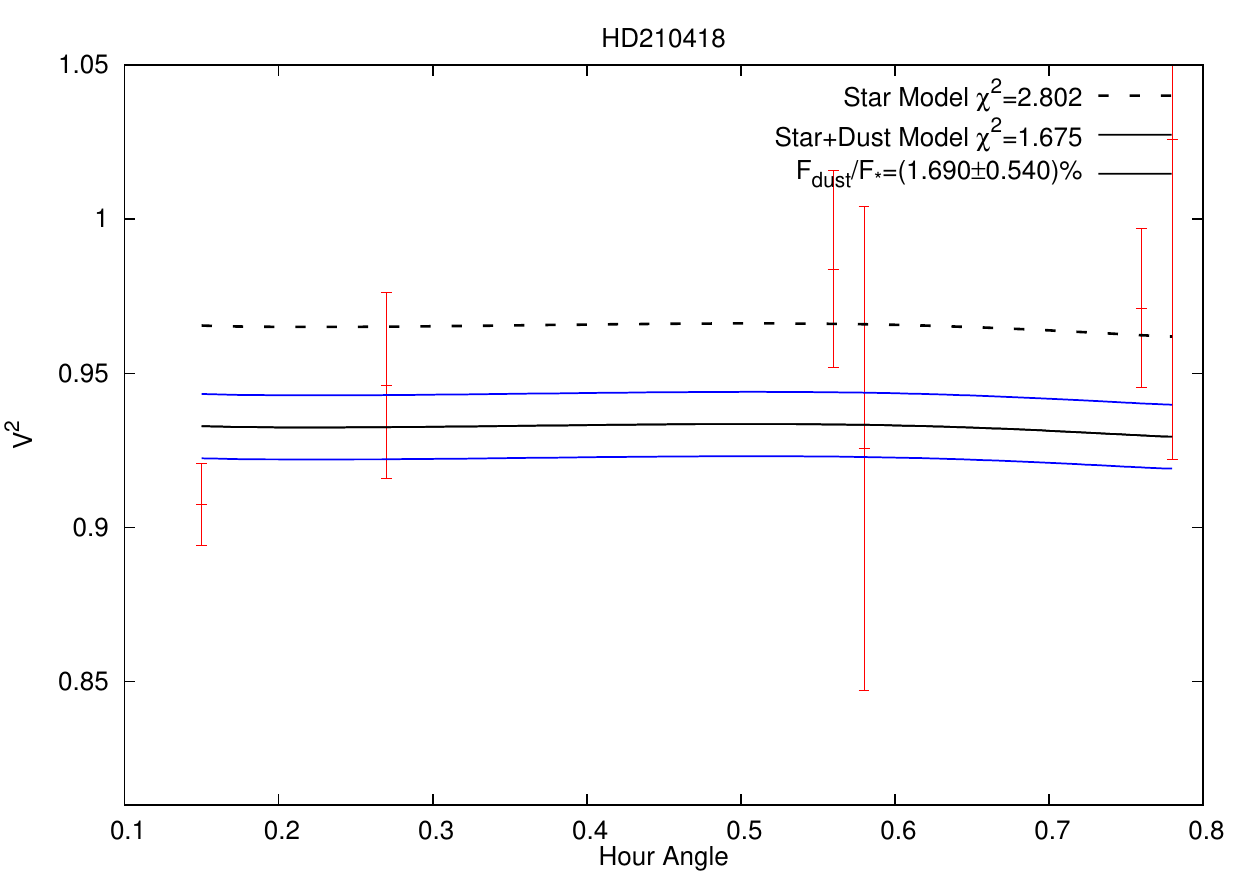} \nonumber
    \end{eqnarray}
  \end{center}
  \caption{\label{detections}Squared visibility as a function of hour angle for the E1-E2 ($65\,\mathrm{m}$) baseline. For these excess star candidates, there is a clear discrepancy between the data and the photosphere model. For both stars the detection significance is above 3$\sigma$ with an acceptable reduced $\tilde{\chi}^2$. However, most of the detection significance of HD210418 is due to a single data point, so more data are needed to confirm this excess.}
\end{figure*}

\textbf{HD222368} is an F-type main sequence star with a cold debris disk as implied by a detected far-infrared excess at $70\mu\mathrm{m}$ with the Spitzer space telescope \citep{trilling_2008}. The angular diameter and linear limb-darkening coefficient were inferred from long baseline observations performed by \citet{taby_2012}. This stellar photosphere model was used to generate the dashed line in Fig. \ref{detections}, which strongly disagrees with the $60\,\mathrm{m}$ baseline data ($\tilde{\chi}^2=4.88$). We report a K-band circumstellar emission of $(1.58\pm0.36)\%$ relative to the star (4.4$\sigma$ detection), with a reduced $\tilde{\chi}^2=1.41$. The data are consistent with a uniform emission over the covered azimuthal angle range ($31^\circ-45^\circ$). This star was also observed with the Keck Interferometric Nuller at $\sim 10\mu\mathrm{m}$, and no excess was detected since the measured null-depth between 8-13$\,\mu$m was $-(0.24\pm1.4)\times 10^{-3}$ \citep{mennesson_2014}.\\

\textbf{HD210418} is an A-type main sequence star that also seems to display a circumstellar excess of $(1.69\pm0.54)\%$ ($\tilde{\chi}^2=1.67$) relative to the star. For this target we also used a stellar photosphere model derived from \citet{taby_2012}. The detection has a lower significance ($3.1\sigma$) and we note that only one data excess point is highly significant (see Fig. \ref{detections}), so  most of the excess comes from a single data point, and more data are required to confirm this detection. This target was observed in the H band with VLTI/PIONIER, and no excess was detected ($-0.43\%\pm 0.29\%$). The JouFLU excess, along with a non-detection using VLTI/PIONIER, allow us to make a rough estimate of the maximum dust temperature. To estimate this temperature, we assume that the excess is due to dust emitting as a graybody, and we further assume that dust is not confined to a certain distance to the star. We estimate that $\sim 1000\,\mathrm{K}$ dust, potentially responsible for the $\sim 1.7\%$ excess at $2.1\,\mu\mathrm{m}$, would likely be undetected by VLTI/PIONIER, since it would correspond to a $\sim 0.3\%$ excess at $1.6\,\mu\mathrm{m}$.

\subsection{Survey excess stars with evidence for binarity: HD33111, HD182640, HD162003, HD5448} \label{binary_sec}

\begin{figure*}
  \begin{center}
    \begin{eqnarray}
        \includegraphics[scale=0.7]{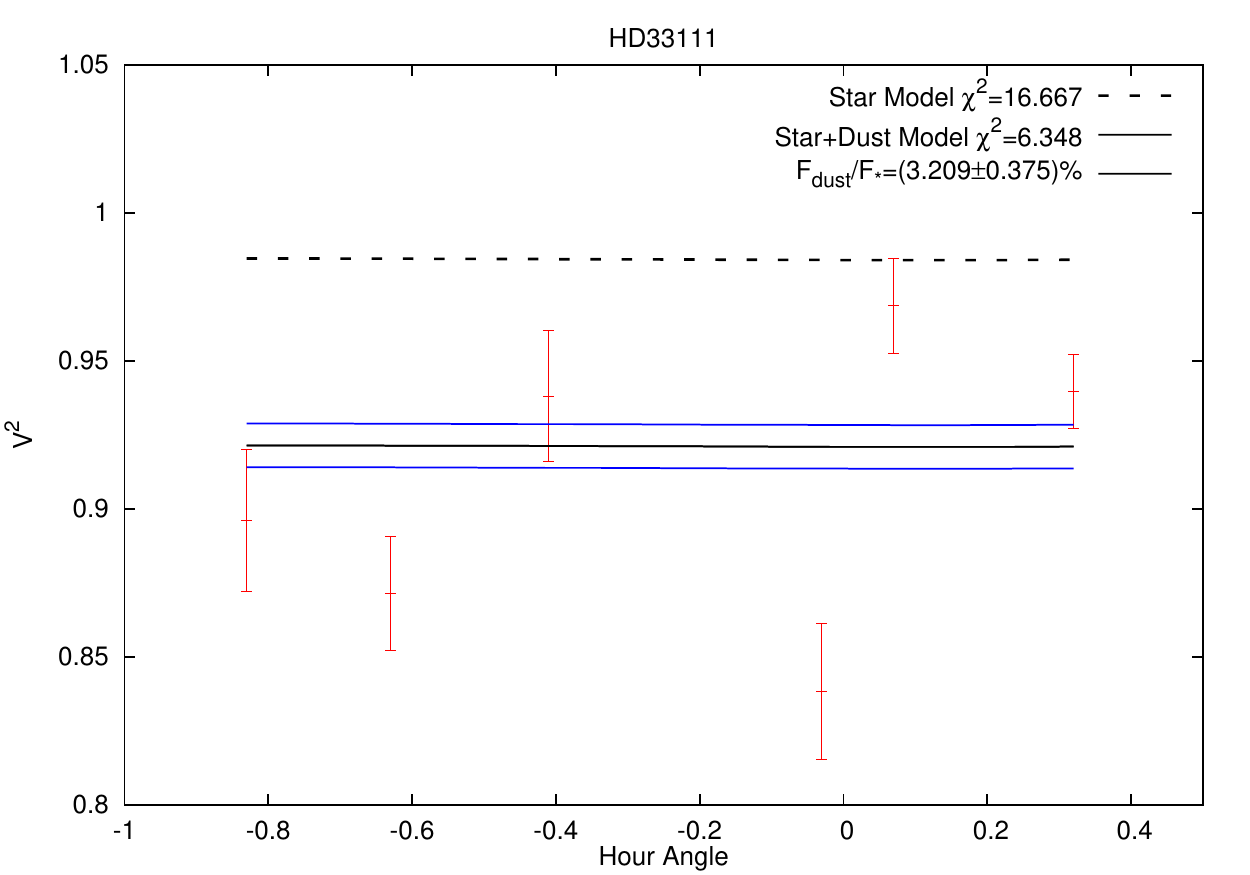} & \includegraphics[scale=0.7]{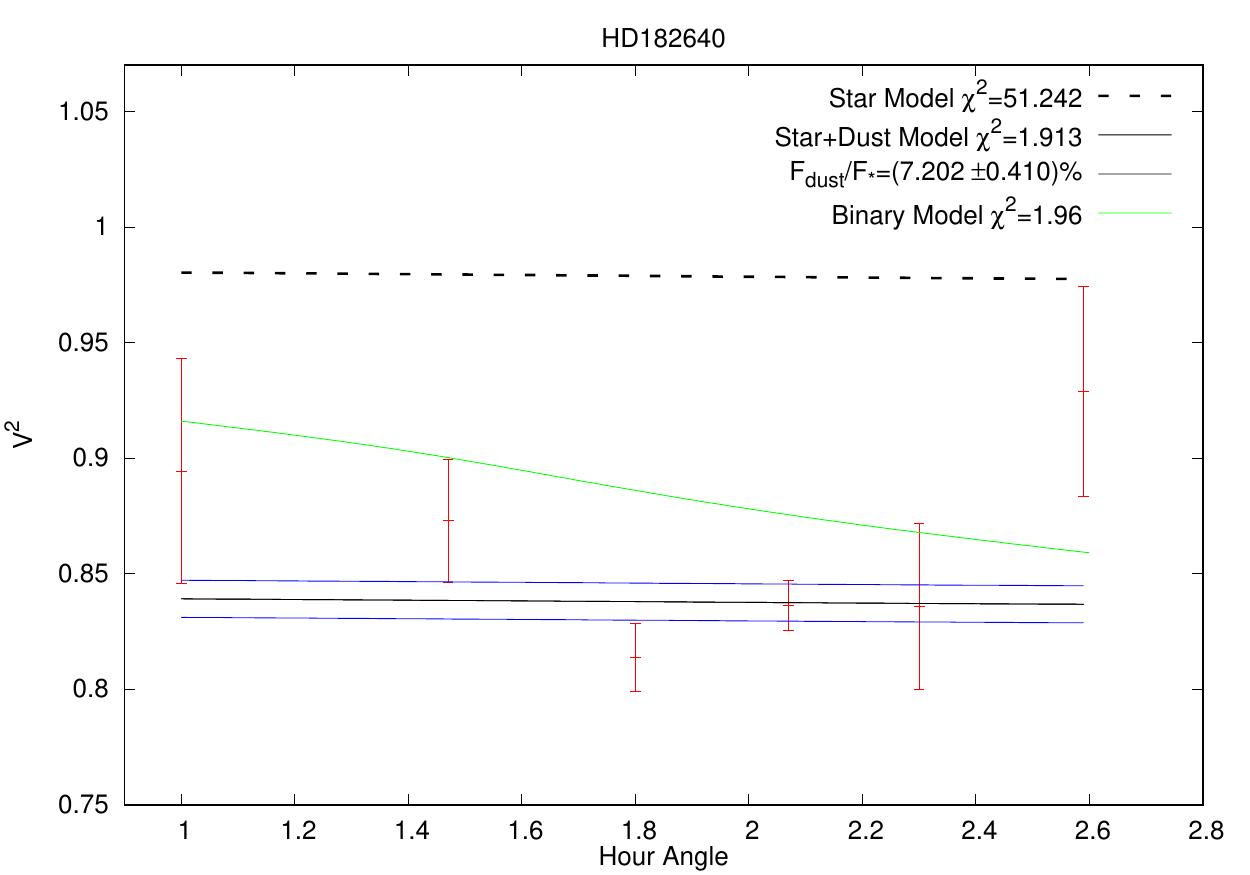} \nonumber \\
        \includegraphics[scale=0.7]{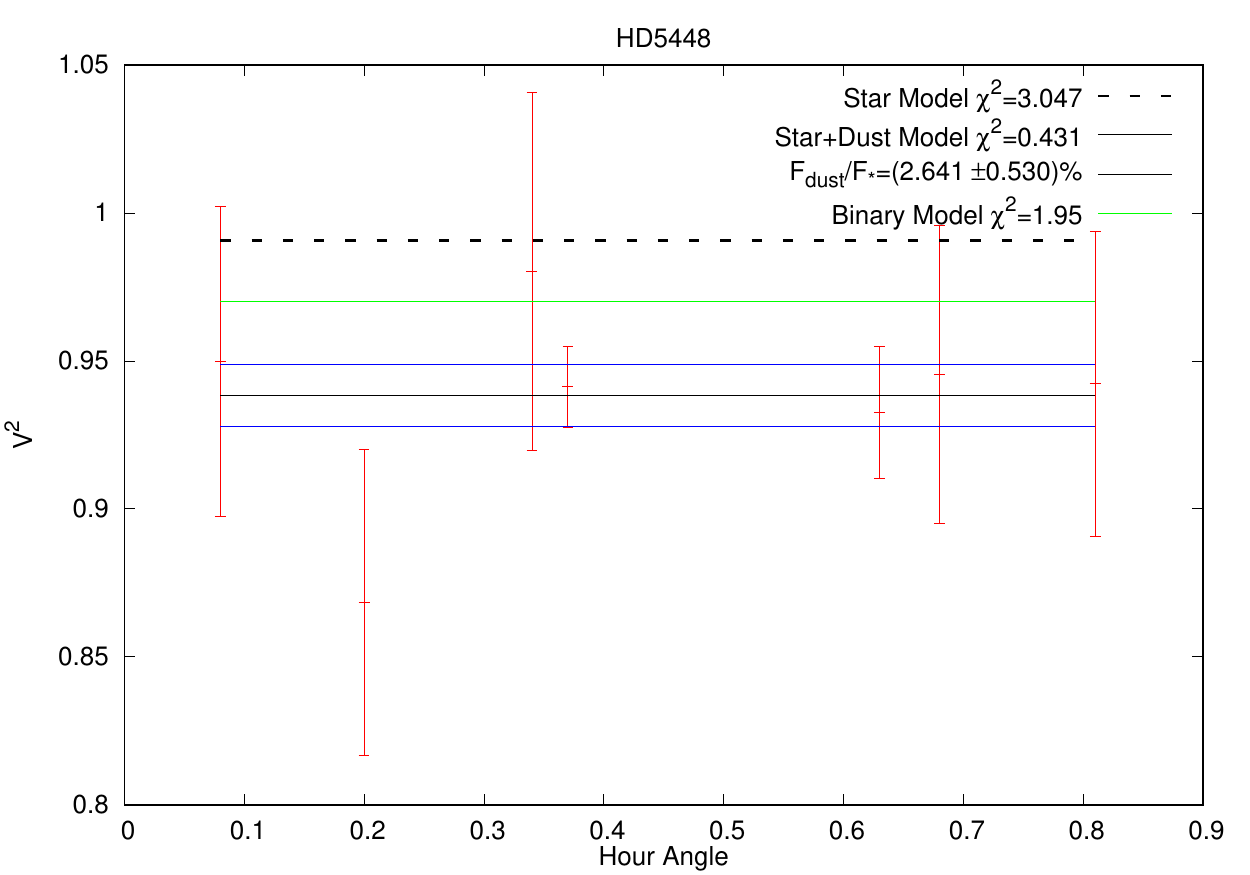} & \includegraphics[scale=0.7]{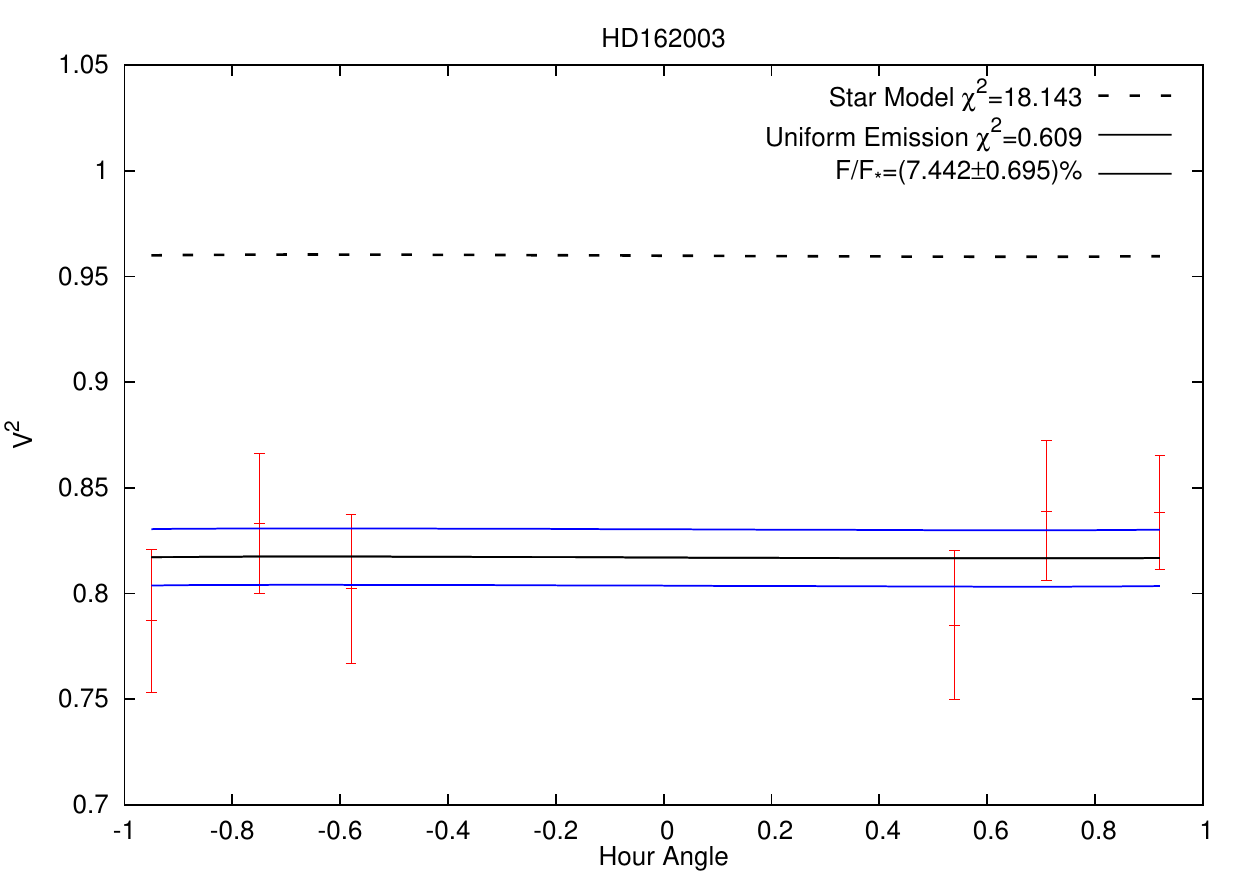}  \nonumber
    \end{eqnarray}
  \end{center}
  \caption{\label{detections2}Squared visibility as a function of hour angle. The S1-S2 ($33\,\mathrm{m}$) baseline was used for observing HD33111, HD182640, and HD5448, and the E1-E2 baseline (65$\,\mathrm{m}$) was used for HD1620003. All stars shown here display a highly significant K-band excesses that we attribute to binarity. See Section \ref{binary_sec}.
}
\end{figure*}

\textbf{HD33111} is an A-type sub-giant displaying a highly significant excess of $(3.21\pm 0.37)\%$. However, as shown in Fig. \ref{detections2}, the high reduced $\tilde{\chi}^2$ of 6.34 indicates a strong disagreement with the pole-on dust model adopted here (Eq. \ref{fcse}), that is, emission is not uniform in the azimuthal angle range of $\sim 82^{\circ}-102^{\circ}$. In view of the large visibility fluctuations over small ($\sim 10^{\circ}$) changes in azimuthal angles, we do not attribute the excess to uniform circumstellar emission, but possibly to binarity. Instead of reporting an excess level for suspected binaries, it is more relevant to report a visibility deficit, which depends on the separation vector at the time of observation, but is more directly related to the brightness ratio between binary components. The visibility deficit for this object is  $1-V/V_{\star}=(3.26\pm 0.38)\%$. However, we note that this target was observed with the VLTI/PIONIER in the H band by \citet{ertel_2014}, who did not report an excess for this target. It is possible that the faint companion was outside the PIONIER FoV during the observations (2012-12-17), or that it was too close to the main star to be resolved by their observations.\\

\textbf{HD182640} is an F-type sub-giant of mass $1.65M_{\odot}$ for which we detect a highly significant visibility deficit of $1-V/V_{\star}=(7.5\pm 0.4)\%$. This star is part of a spectroscopic binary system with a semi-major axis of $\sim 50\,\mathrm{mas}$, well within the FOV of JouFLU. Since this is a known binary, we mainly use these data to check that our measured visibilities are compatible with a binary model. The stars of this system have masses of  $0.67M_{\odot}$ and $1.65M_{\odot}$ \citep{fuhrmann_2008}. Assuming temperatures of $7200\,$K, typical for an F0IV, and $4300\,$K for the faint companion, we estimate the K-band flux ratio to be $\sim 6.6\%$. Using this estimate of the contrast ratio, we make a detailed analysis of the compatibility of the data with a binary model, by using the orbital solution obtained with HIPPARCOS \citep{hipparcos_1997}, as well as the ASPRO software \citep{aspro}, we simulate the squared visibilities at the time of our observations, shown in Fig. \ref{detections2}. The binary orbital parameters and the data are in reasonable agreement in view of the reduced $\tilde{\chi}^2=1.96$.\\
 
We also report a significant K-band visibility deficit for \textbf{HD5448} (Fig. \ref{detections2}, bottom-left panel), namely $1-V/V_{\star}=(3.0\pm 0.5)\%$, using the S1-S2 ($\sim 33\,\mathrm{m}$) baseline. In view of the recent results obtained by \citet{fabien_2014} and \citet{rachel_2016}, using the MIRC beam-combiner at the CHARA array, this visibility deficit is not likely related to hot dust, but rather due to a faint companion. \citet{rachel_2016} report an H-band flux ratio of $(1.25\pm0.3)\%$ using closure phase measurements, and their orbital parameters correspond to a highly eccentric orbit with a $\sim 47\,\mathrm{mas}$ semimajor axis approximately oriented in the north-south direction, that is, closely parallel to the S1-S2 baseline used in the JouFLU observations reported here. The mass of HD5448A is $\sim 2M_{\odot}$, and the orbital parameters measured  by \citet{rachel_2016} allow estimating the mass of the faint companion to $\sim 0.81M_{\odot}$. With these mass estimates, we assume temperatures of $8200\,$K and $5300\,$K, which allow us to estimate the K-band flux ratio of $\sim (1.4\pm0.5)\%$, which is in turn $\sim 2.3\sigma$ lower than our measurement. If we use the orbital parameters found by \citet{rachel_2016}, and the ASPRO software to simulate the squared visibilities at the time of observation, the best fit binary model predicts an essentially constant squared visibility of $\sim 0.97$ at the time of observations, which is in acceptable agreement with the data as evidenced by a reduced $\tilde{\chi}^2=1.9$.\\

One of the highest detected K-band excess of all the targets is for \textbf{HD162003A} as shown in Fig. \ref{detections2} (botton-right panel). We report a visibility deficit of $1-V/V_{\star}=(7.7\pm 0.75)\%$ at the E1-E2 ($\sim 65\,\mathrm{m}$) baseline, where the photosphere model was derived from long-baseline measurements obtained by \citet{taby_2012}. This object is part of a visual binary star with an angular separation of $\sim 30.1''$, which is well beyond the FOV of JouFLU ($\sim 0.5''$ in radius), so we are not detecting HD162003B. HD162003A, with a mass of $1.43M_{\odot}$, does have a long period radial velocity trend as reported by  \citet{toyota_2009}, corresponding to a low mass M-dwarf of $(0.526\pm 0.005)M_{\odot}$ \citep{endl_2016}, but a complete binary model cannot yet be determined from observations. We can use the stellar masses to  assume the temperatures of the binary components to be $6500\,$K and $3700\,$K,  to estimate the K-band flux ratio of $\sim 6.2\%$. This estimated K-band ratio is compatible with the measured visibility deficit, so we are possibly directly detecting this unseen companion for the first time.\\

\section{Results of the follow-up observations of the initial FLUOR survey} \label{follow_up}

All of the stars that were part of the follow-up program had been reported as hot dust candidates, with the exception of HD9826 and HD142860. The follow-up measured excess levels are shown in Table \ref{follow_up_table}. Most excess levels are compatible with those reported by \cite{absil_2013}, as shown in the ninth column of Table \ref{follow_up_table}, but four stars differ significantly: namely  HD9826, HD142091, 187642, and HD173667,  which is a newly suspected binary as we discuss below. We also note that the typical uncertainties for the follow-up program, shown in Table \ref{follow_up_table}, are larger than those published by \citet{absil_2013}, by about a factor of between approximately two and three, for reasons that are currently under investigation\footnote{Some polarization mismatches were detected in the JouFLU beam combiner back in 2013 \citep{nic_thesis}. They caused a significant (x2.5) decrease in the instrumental point source visibility, and were also suspected to amplify the instrument sensitivity to varying observing conditions (e.g temperature or zenith angle). In June of 2016, we installed Lithium Niobate plates that compensate for differential birefringence, as described by \citet{lazareff_2012}, which resulted in a notable increase in the raw visibility values, from $\sim 0.3$ in 2013, to $\sim 0.6$ in 2016.}. Therefore, of the nine stars that were reported as displaying a circumstellar excess by \citet{absil_2013}, we only re-detect four as having a significant circumstellar excess attributed to dust. We discuss some results for individual targets below.\\

The object \textbf{HD9826} ($\upsilon$ And) is known to host four Jovian (RV) planets as close as $\sim 0.06\,$AU ($\sim 4.4\,$mas). $\upsilon$ And does have known stellar companions \citep{lowrance_2002}, but these are well outside of the FOV at 55, 114, and 287''. \citet{absil_2013} reported a circumstellar excess for this star of $(0.53\pm 0.17)\%$ (S1-S2, 33m baseline), which has a significance of $3.1\sigma$, but did not report it as a hot dust candidate because the excess level significance remained marginal. We report a much higher K-band circumstellar excess of $(3.62\pm 0.61)\%$ when using the same $\sim 33\,\mathrm{m}$ baseline as shown in Fig. \ref{follow_up}. The data are consistent with no evidence for a significant  azimuthal variation in the range $\sim 75^{\circ} - 95^{\circ}$, for both the JouFLU and the previous FLUOR observations. Since the position angles of the FLUOR and JouFLU observations cover essentially the same azimuthal range, it is much less likely that the excess varibility is due to the visibility signature of an inclined disk. If this excess is indeed due to exozodiacal dust emission, we are therefore detecting variability\footnote{Variability on the exozodi light level has also been detected on at least one other object (HD 7788) by \citet{ertel_2016}.} in the exozodi light level of this object. 

This star has been observed extensively by RV instruments (e.g., \citet{buttler_1999}), and a bright stellar companion contributing several percent of the infrared flux within the 0.5'' JouFLU FOV would have been likely detected, unless its orbit is seen very close to pole-on and in a very different plane than the planets, which is unlikely. We also exclude the possiblity of binarity based on other interferometric campaigns: an extensive search for close companions on HD9826 was made with the CLASSIC beam combiner \citep{baines_2008} in the K band, and with the MIRC beam combiner (H-K bands) \citep{zhao_2008} at the CHARA array, and none found any stellar companions within 1''. MIRC in particular would have  detected a close companion contributing $\sim 1\%$ of the flux, and located anywhere between a few milliarcseconds to 1'' away from the star.

\begin{table*} 
\caption{\label{follow_up_table}Circumstellar emission for the follow-up targets of \citet{absil_2013}.  The highlighted boxes in the sixth column correspond to statistically significant excesses: red boxes indicate that the excess is statistically significant, and also consistent with the circumstellar dust model of Eq. \ref{fcse}. We highlight in yellow the excess of HD 173667 because it is likely due to binarity. We also report the FLUOR excess levels reported by \citet{absil_2013} (Col. 9), and the change in circumstellar emission $\Delta f_{cse} = f_{cse,\mathrm{JouFLU}}-f_{cse,\mathrm{FLUOR}}$ (Col. 10), the difference between the results obtained here, and those published by \citet{absil_2013}. We find discrepant results for four stars, which are highlighted in the last column. See text for details.}

\scalebox{1.0}{
\begin{tabular}{ccccclcccc}
\hline\hline
HD & Name & Type & K & Baseline & $f_{cse}$(\%) \tiny{JouFLU}  & $\tilde{\chi^2}_{cse}$  &  $\tilde{\chi^2}_{\star}$ &  $f_{cse}$(\%) \tiny{FLUOR} & $\Delta f_{cse}$(\%)\\ \hline
9826 & $\upsilon$ And & F9V & 2.84 & S1-S2 & \cellcolor{red!25} $3.62\pm0.61$ & 1.26 & 4.42   & $0.53\pm 0.17$    &  \cellcolor{red!25} $3.09\pm0.63$\\ \hline

40136 &  $\eta$ Lep & F2V & 2.91 & S1-S2 & $0.45\pm0.47$ & 2.03 & 1.8   & $0.89\pm0.21$  &  $0.44\pm0.52$ \\ \hline 

102647 & $\beta$ Leo & A3Va & 1.91 & S1-S2 & $1.09\pm0.52$ & 0.90 & 1.44 & $0.94 \pm 0.26$  & $0.15\pm 0.59$\\ \hline 

131156 &  $\xi$ boo & G7V & 1.97 & S1-S2 & $-0.32\pm0.42$ & 1.40 & 1.29  & $0.74 \pm 0.20$   & $-1.06\pm0.47$\\ \hline 

142091 & $\kappa$ CrB & K1IV & 2.49 & \tiny{S1-S2, E1-E2} & \cellcolor{red!25} $3.4\pm0.52$ & 1.85 & 6.54 & $1.18 \pm 0.20$ & \cellcolor{red!25}$2.22\pm0.56$\\ \hline 

142860 & $\gamma$ Ser & F6IV & 2.62 & E1-E2 & $1.09\pm0.78$ & 1.02 & 1.17 & $-0.06 \pm 0.27$  &  $1.15\pm 0.94$ \\ \hline

173667 & 110 Her & F6V & 3.05 & S1-S2 & \cellcolor{yellow!25} $2.34\pm0.37$ & 10.66 & 9.24  & $0.94\pm0.25$ & \cellcolor{red!25} $1.40\pm 0.44$\\ \hline 

177724 &  $\zeta$ Aql & A0Vn & 2.9 & E1-E2 & \cellcolor{red!25} $1.23\pm0.38$ & 1.32 & 2.45 & $1.69\pm0.27$ & $-0.46\pm 0.47$\\ \hline 

185144 &  $\sigma$ Dra & G9V & 2.81 & S1-S2 & $-1.11\pm1.0$ & 1.88 & 1.78 & $0.15\pm 0.17$ & $-1.26\pm1.01$ \\ \hline 

187642 &  $\alpha$ Aql & A7V & 0.22 & S1-S2 & \cellcolor{red!25} $6.11\pm0.74$ & 2.76 & 27.33 & $3.07\pm0.24$  & \cellcolor{red!25} $3.05\pm 0.78$\\ \hline 

203280 & $\alpha$ Cep & A7IV & 1.96 & E1-E2 & $-0.14\pm0.78$ & 0.74 & 0.64 & $0.87\pm 0.18$ & $-1.01\pm 0.8$\\ \hline 

\end{tabular}
}

\end{table*}

In addition, speckle observations on HD9826 were recently made with NESSI (NN-EXPLORE Exoplanet \& Stellar Speckle Imager) at the WIYN observatory in October of 2016. This instrument is a dual-channel (562 and 832 nm) speckle camera that uses two Andor EMCCD cameras based off of the previous DSSI (Differential Speckle Survey Instrument) for diffraction-limited imaging (see \citealt{horch_2011}). We use these data to constrain the possibility that the excess flux thought to be due to exozodiacal dust is the result of a faint companion. The wider FOV of 2.3'' from the speckle observations rules out any such companion at the edge of the JouFLU FOV. An upper limit of the flux ratio at 832$\,\mathrm{nm}$ of a companion located 0.1'' away was found to be 1.6\%, which allows us to exclude any hypothetical companion of earlier spectral type than M8V/M9V (T $\geq$ 2500$\,$K). This contrast limit increases with greater separation of the hypothetical companion, reaching 0.4\% (T=2100$\,$K) at the extent of the JouFLU FOV and 0.06\% (T=1700$\,$K) at the extent of the speckle FOV. Even in the minimal separation case the speckle observations exclude companions of earlier spectral type than M8V/M9V (T $\geq$ 2500 K), and provides supporting evidence that the observed JouFLU excess is not due to binarity.\\

\begin{figure*}
  \begin{center}
    \begin{eqnarray}
        \includegraphics[scale=0.7]{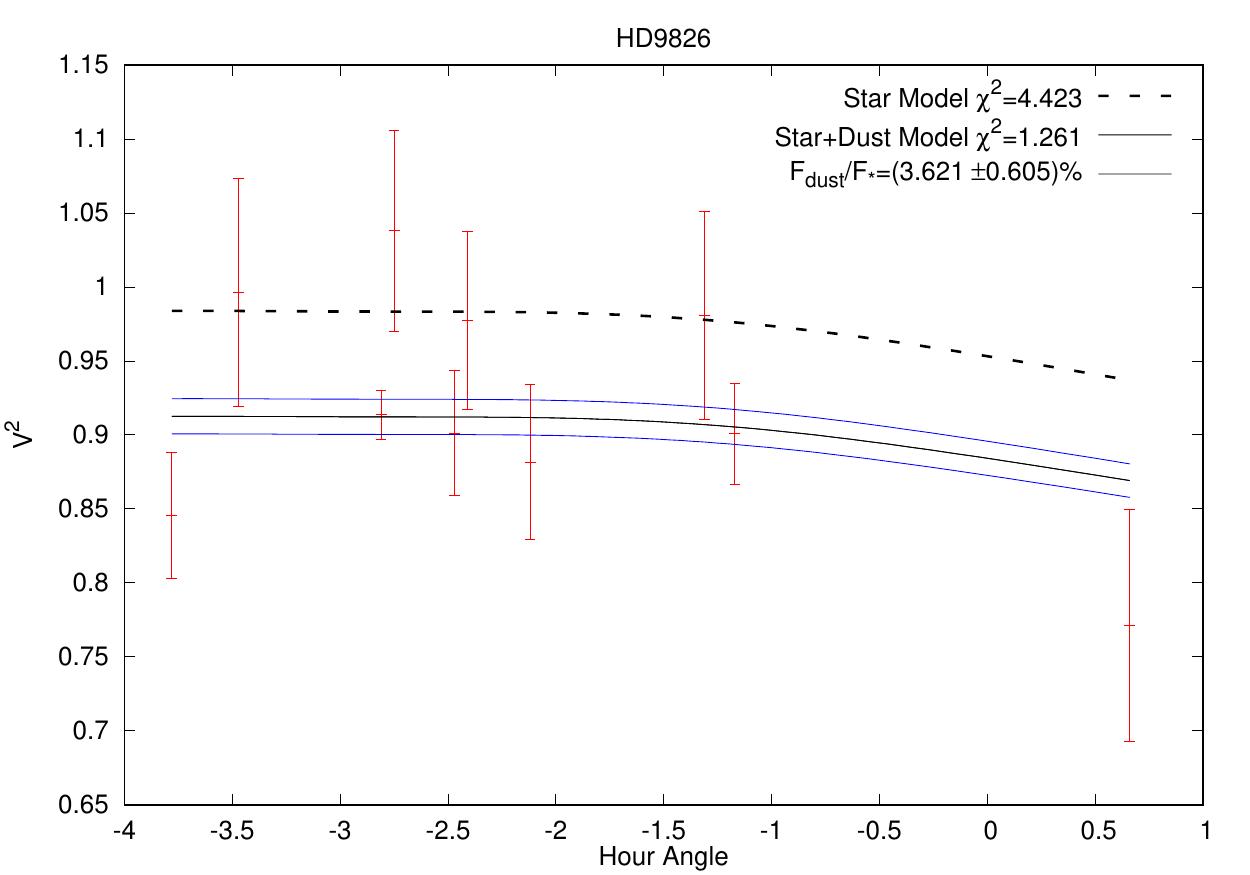} & \includegraphics[scale=0.7]{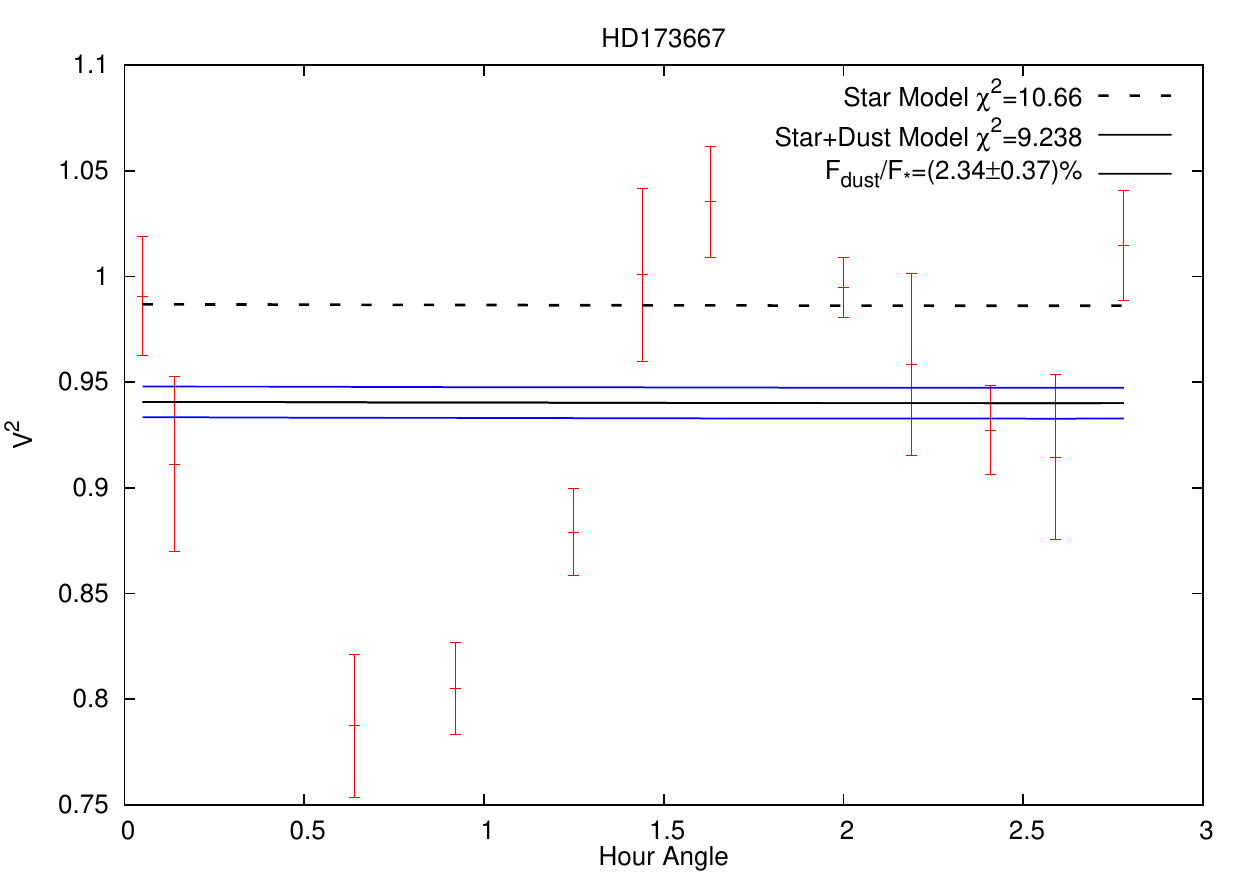} \nonumber
    \end{eqnarray}
  \end{center}
  \caption{\label{follow_up} Measured K-band excess for HD9826 (left panel) with the S1-S2 (33m) baseline. Also shown is the binary behavior of HD173667 (right panel).}
\end{figure*}

\begin{figure*}
  \begin{center}
    \begin{eqnarray}
      \includegraphics[scale=0.7]{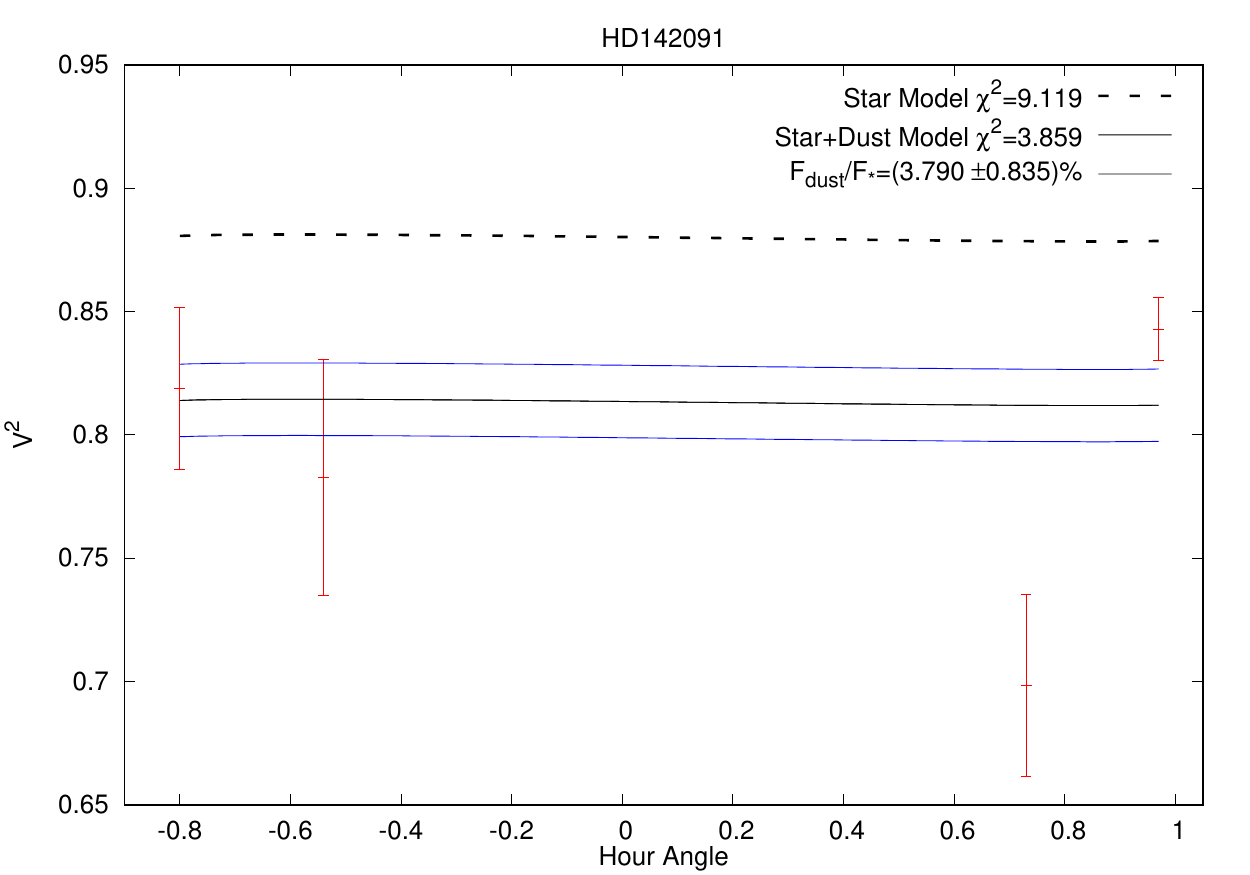} & \includegraphics[scale=0.7]{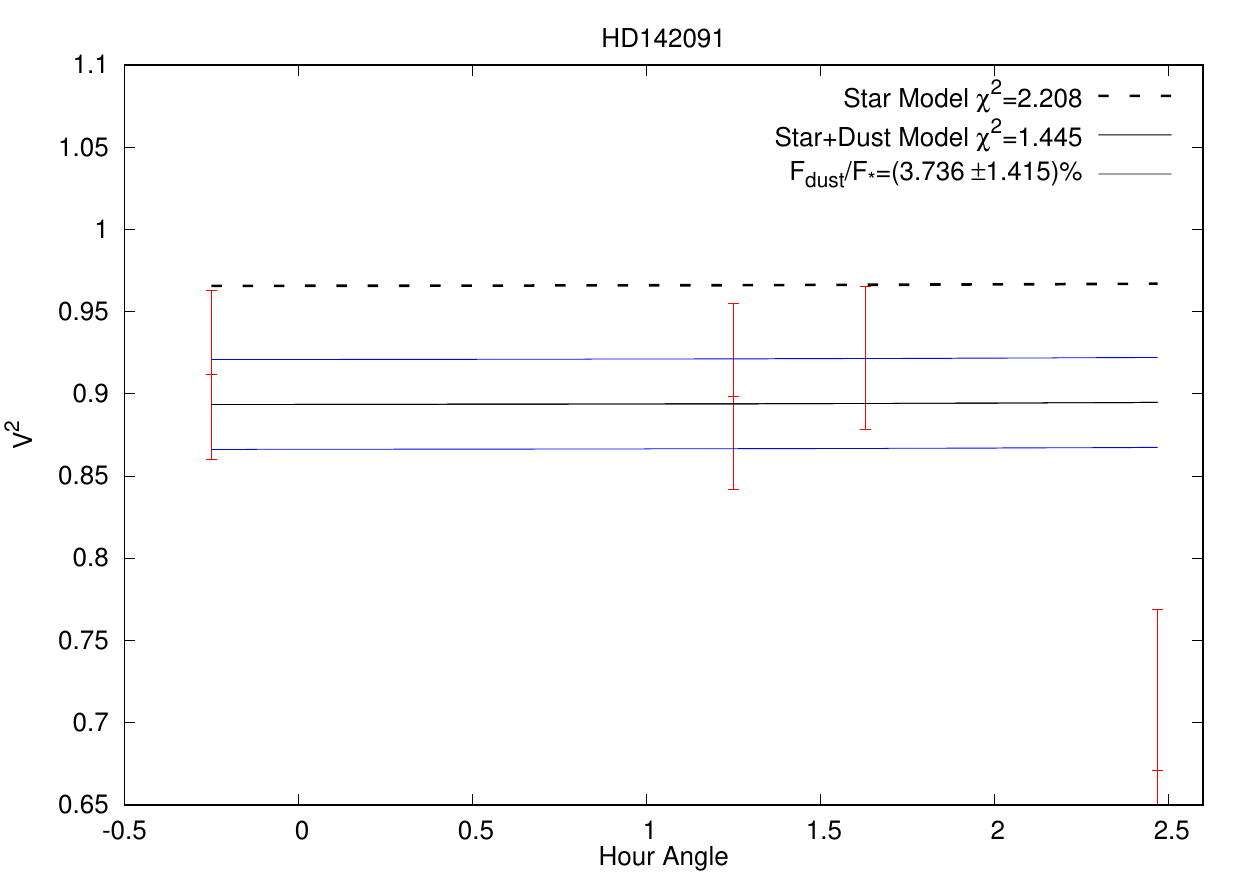} \nonumber
    \end{eqnarray}
  \end{center}
  \caption{\label{kappa_crb} Left Panel: Measured a significant K-band excess for the E1-E2 (63$\,$m) baseline. Right panel: measurements using the S1-S2 (33$\,$m) baseline.}
\end{figure*}

In the case of \textbf{HD142091} ($\kappa$ CrB), we detect an excess of $(3.4\pm0.52)\%$, with no significant azimuthal variation. This detection is also particularly interesting since this star is known to host two exoplanets \citep{johnson_2008}. Moreover, resolved images of a debris disk extending as far as $\sim 300\,\mathrm{AU}$ from the star, were obtained by \citet{bonsor_2013} using Hershel in the far-infrared. \citet{bonsor_2013} also performed high contrast observations in the H band with NIRC2 and the Keck adaptive optics system, and upper limits allow discarding a $\sim 0.06\%$ companion located $\sim 0.2''$ away from the star, with their contrast limit increasing with greater separation of a hypothetical companion. The JouFLU observations were performed at two different baselines, and we measured a $(3.8\pm 0.8)\%$ excess with the E1-E2 (63$\,$m) baseline, and $(3.7\pm 1.4)\%$ with the S1-S2 (33$\,$m) baseline, both measurements consistent with each other, although with some very low visibility outliers that can be seen in Fig. \ref{kappa_crb}. The radial velocity and high contrast observations allow ruling out any stellar companion closer than 0.2'', so the K-band circumstellar emission detected by JouFLU, and previously by \citet{absil_2013}, most likely  comes from hot dust within $\sim 12\,$AU of the star, where at least one of the planets is thought to lie. This star has started to depart from its main sequence, but K-band excess is not likely due to partially resolved atmospheric structures, because we note that in order to explain the K-band excess, the apparent angular diameter would need to be $\sim 20\%$ greater than the adopted value of $(1.54\pm 0.009)\,\mathrm{mas}$ found by \citet{baines_2013}. We see a statistically significant increase of $\sim 2\%$ in the K-band excess level compared to results published by \citet{absil_2013}. However, it is not straightforward to claim a variability in the circumstellar emission level at this point, since the significant JouFLU excess was measured using the E1-E2 ($63\,$m) baseline, whereas the excess reported by \citet{absil_2013} relied on S1-S2 (33$\,$m) baseline data. \\

\begin{figure}
  \begin{center}
    \includegraphics[scale=0.72]{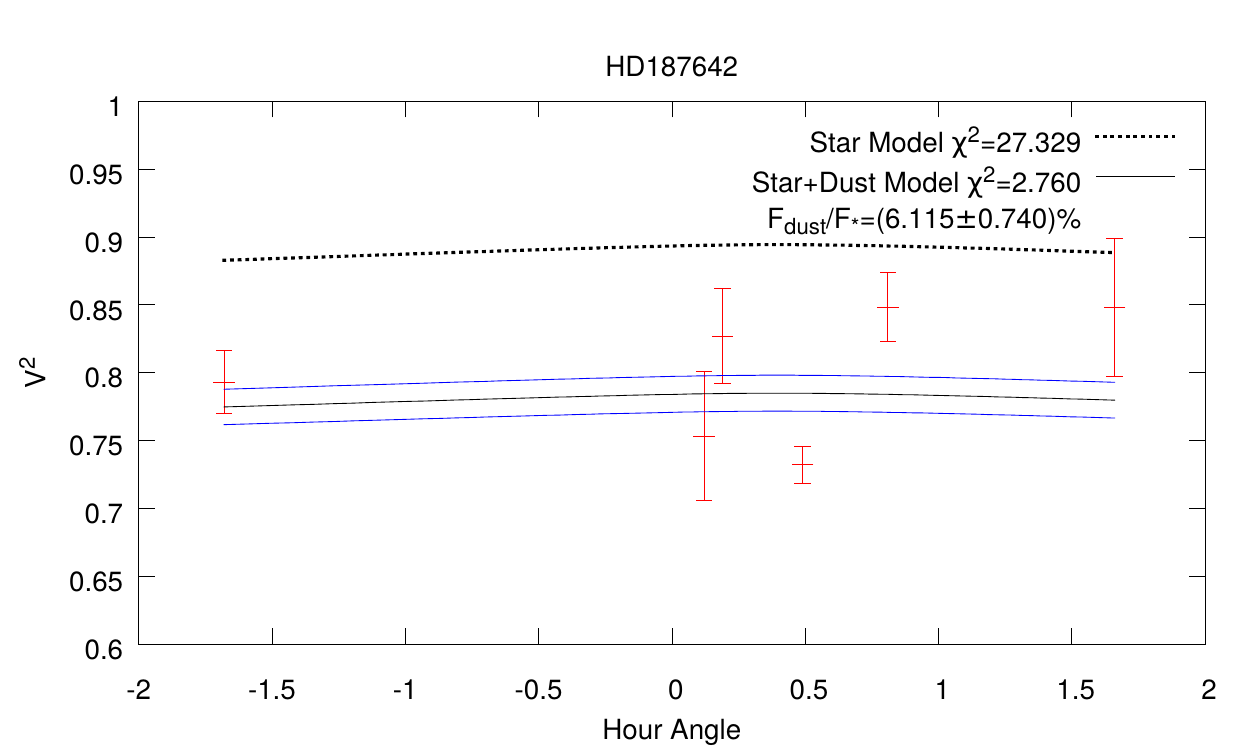} 
  \end{center}
  \caption{\label{altair} K-band excess of $6.1\%$ was detected for Altair, using the S1-S2 ($33\,$m) baseline.}
\end{figure}

\textbf{HD187642} (Altair) is displaying significant variability in its circumstellar emission. This star displayed the highest K-band excess reported in the \citet{absil_2013} sample, namely $(3.07\pm 0.24)\%$, using the S1-S2 ($33\,$m) baseline. Using the same baseline, we report a very high excess of $(6.11\pm 0.74)\%$ as shown in Fig. \ref{altair}, as infered by the visibility deficit relative to the stellar model derived by \citet{monnier_2007}. We can exclude a faint companion based on extensive observations with CHARA/MIRC in the H band, which resulted in detailed images of the surface of this star \citep{monnier_2007}. The closure-phase signals of the MIRC observations have been shown to be capable of detecting faint  ($\sim 1\%$) \cite{rachel_2016} and close ($<0.5''$) stellar companions, and would have easily detected a companion responsible for the K-band excess.\\

Another case where we see a discrepancy is \textbf{HD173667} (110 Her), which was previously reported to have an excess of $(0.94\pm 0.25)\%$  by \citet{absil_2013}. We detect a higher excess of $(2.34\pm 0.37)\%$ using the same baseline (S1-S2, 33$\,$m). The data display large ($\sim 20\%$) fluctuations in the squared visibility as shown in Fig. \ref{follow_up}, which result in a very poor fit to both the photosphere model, derived from parameters obtained by \citet{taby_2012}, and the star and dust model. The large modulation of the squared visibility measurements obtained versus hour-angle suggests that the detected excess is likely due to  a stellar companion. However, a nearly edge on disk with very bright structures can not be entirely discarded given our limited observations. \\

Interestingly, we can discard the possibility that any of these changes are due to the new data reduction software (DRS): we used the new DRS to re-analyze some of the data previously obtained with FLUOR, and found excess levels and uncertainties compatible with those published by \citet{absil_2013}.

\section{Overall statistical analysis} \label{statistical_analysis}

For the analysis presented below we did not include JouFLU survey targets which show strong evidence of binarity (HD33111, HD162003, HD182640, HD5448). This conservative approach leads to two new JouFLU excess detections imputable to dust among the 29 - presumably- single stars observed with the JouFLU survey. Similarly, for the FLUOR sample \citep{absil_2013}, we did not include the confirmed binary HD211336 \citep{mawet_2011}, resulting in 12/41 FLUOR detections.

\subsection{Significance distribution of the JouFLU survey}

\begin{figure*}
  \begin{center}
    \begin{eqnarray}
      \includegraphics[scale=0.45]{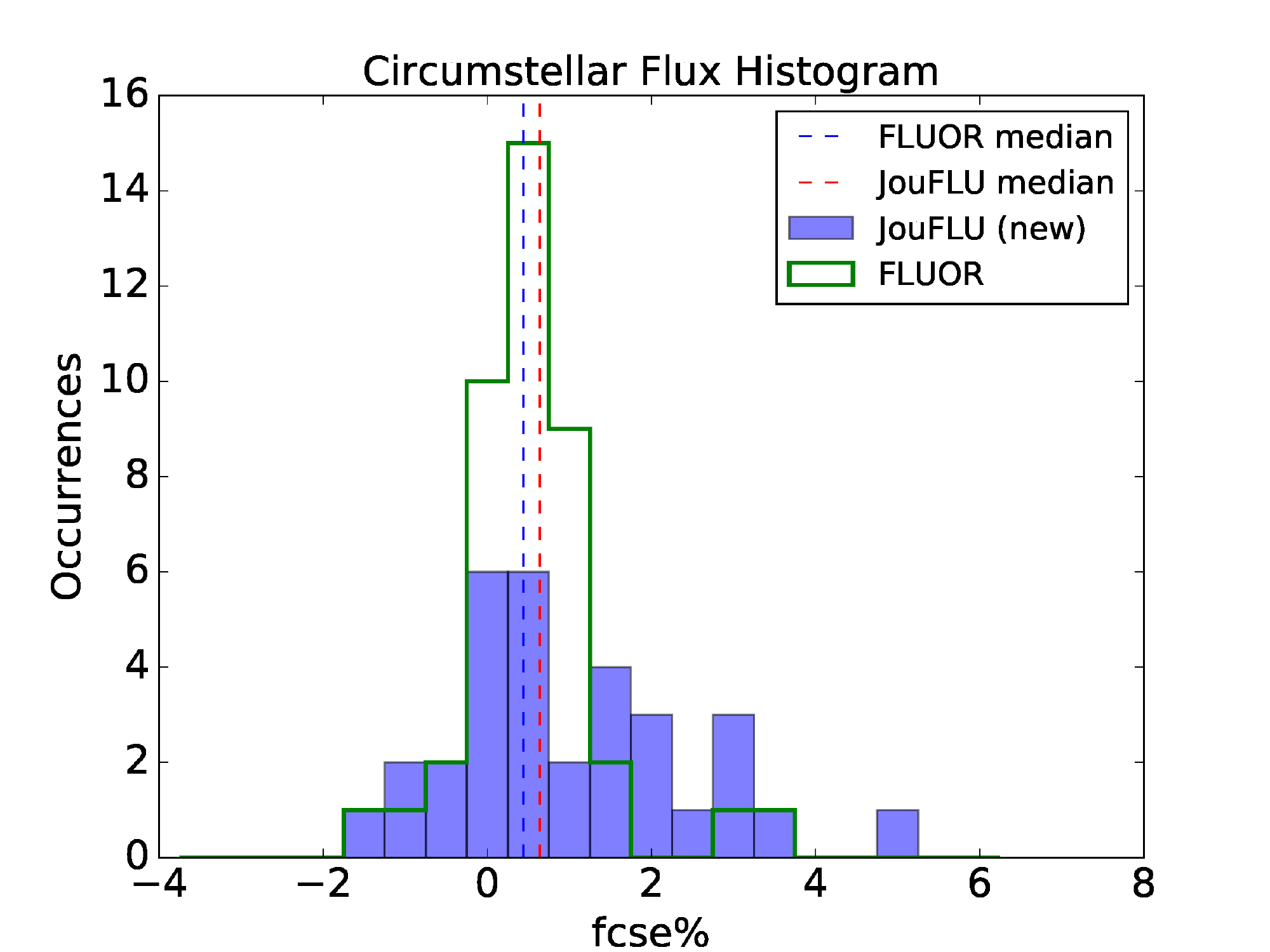} & \includegraphics[scale=0.45]{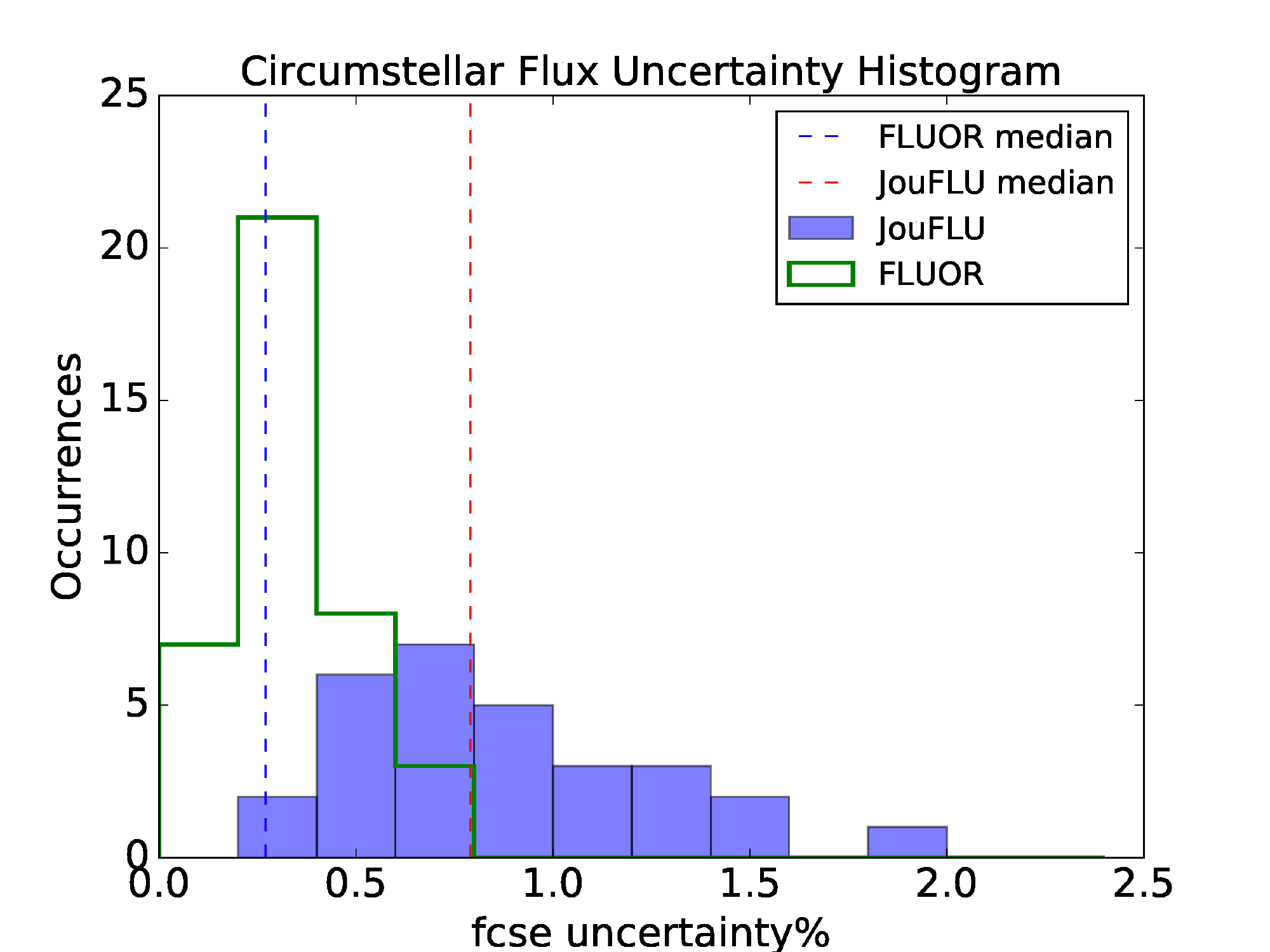} \nonumber
    \end{eqnarray}
  \end{center}
  \caption{\label{csf_histo_old_new}Measured circumstellar emission (left panel) and its $1\sigma$ uncertainty (right panel) for the new JouFLU targets (blue), and the FLUOR circumstellar emission values reported by \citet{absil_2013} (green) }
\end{figure*}

First we examine the circumstellar excess levels, and their uncertainties, for the new stars added through the CHARA/JouFLU 2013-2016 survey. In the left panel of Fig. \ref{csf_histo_old_new} we compare the measured circumstellar emission levels measured in the \citet{absil_2013} FLUOR survey, with the emission levels measured by the JouFLU survey (this work). The weighted mean of the measured circumstellar emission ($f_{\mathrm{cse}}$) of the FLUOR targets is $(0.06\pm 0.05)\%$ and $(0.36\pm 0.12)\%$ for the JouFLU targets. We note that there is a significant positive bias on the measured excess levels for the JouFLU survey with a significance of $\sim 0.36/0.12=3.0$, which may be evidence of an underlying population of undetected excess stars. As shown in the right panel of Fig. \ref{csf_histo_old_new}, there is a notable increase in the median uncertainty: from $0.27\%$ for FLUOR to $0.78\%$ for JouFLU. To some extent, this higher uncertainty is expected, given that stars in the JouFLU sample are typically fainter than in the FLUOR survey by a magnitude of approximately one.

\begin{figure}
  \begin{center}
    \includegraphics[scale=0.45]{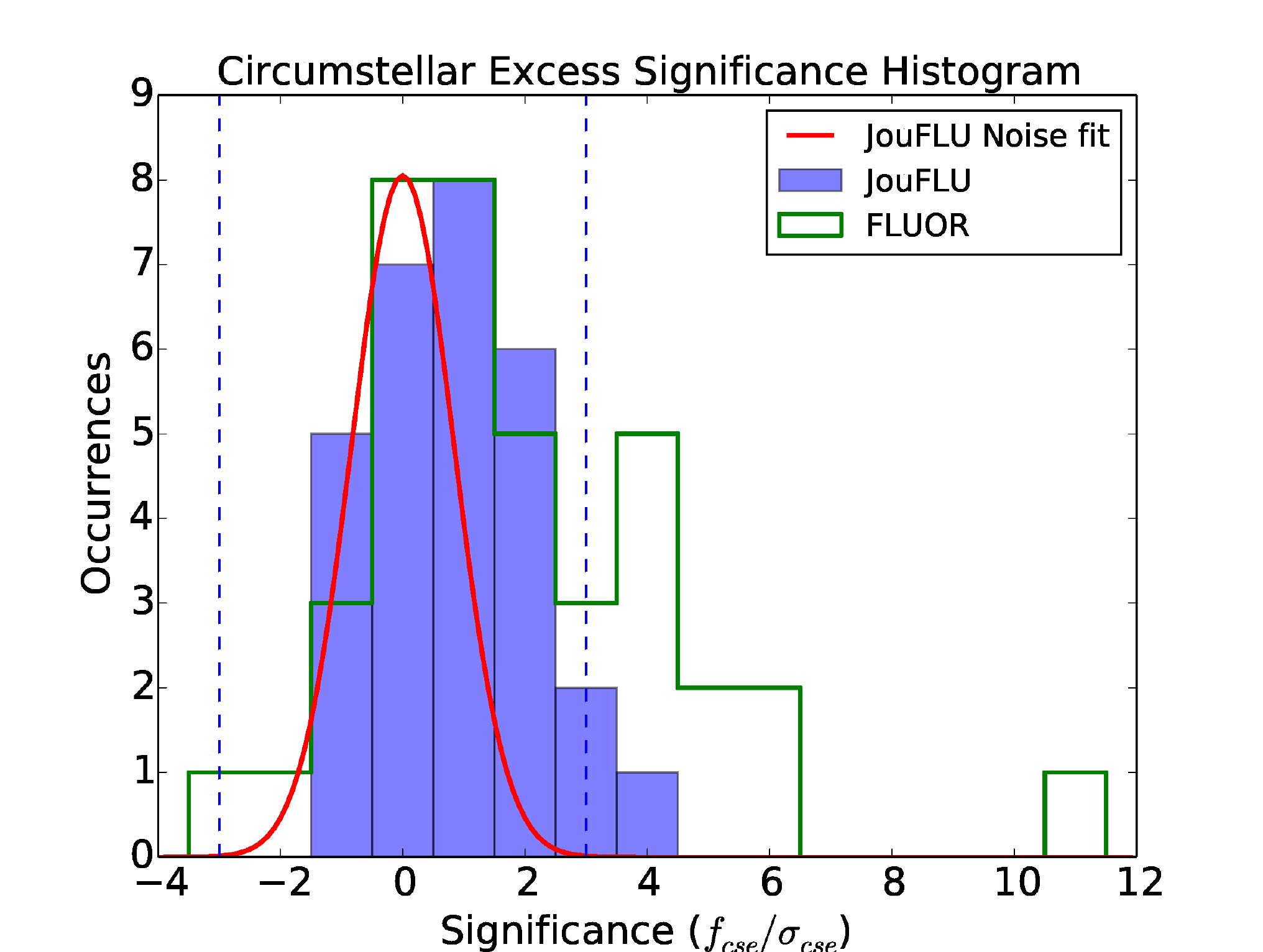}
  \end{center}
  \caption{\label{sig_histo_old_new} Significance histogram for the JouFLU survey  shown in blue, and for the FLUOR survey \citep{absil_2013} shown in green. The red curve is an estimate of the instrumental noise of JouFLU, which is computed by fitting a Gaussian to a symmetrized JouFLU distribution around zero. The difference of between the instrumental noise and the JouFLU significance distribution allows to estimate the number of undetected excesses, which is around $\sim 9$.  }
\end{figure}

Now we compare the significance distributions of the JouFLU and FLUOR samples separately, where by significance we mean the ratio of the measured excess level to its measured uncertainty ($f_{\mathrm{cse}}/\sigma_{\mathrm{cse}}$). In Fig. \ref{sig_histo_old_new}, we show the significance histogram reported by \citet{absil_2013} and compare it with the signficance histogram derived for the JouFLU targets.

Both significance distributions are biased toward positive excess. In particular, the distribution of the JouFLU survey has a median of 0.73. As stated above, this bias may be evidence of a population of undetected excess stars that we can further quantify below. By assuming that the negative significance bins are due to Gaussian noise, we can symmetrize the distribution around zero significance to estimate the instrumental noise distribution. We then fit a Gaussian to this JouFLU instrumental noise distribution, taking into account the finite bin-width, and we find a best-fit standard deviation of 0.84. This is fairly close to unity, although a bit smaller, which may be expected from a limited number (12) of independent realizations of $f_{\mathrm{cse}}/\sigma_{\mathrm{cse}}$. This may also suggest that our error bars are slightly overestimated.

Figure \ref{sig_histo_old_new} shows the instrumental noise distribution and the excess significance measured for the JouFLU targets.  As already discussed in Section \ref{survey_detections}, two of the presumably single stars retained for this analysis do show a bona fide K-band excess, meaning that they are  detected at the $3\,\sigma$ level or higher.  But there are also many more targets with excess significance levels between 1 and $3\,\sigma$ than predicted by instrumental noise alone. In addition to the two stars with significant excesses, this suggests that approximately ten more targets have excesses close to the JouFLU detection limit. We just do not know which ones.

Regarding our finding of a positive mean value of the circumstellar emission: it is possible that calibration biases could result in a decrease of the transfer function, either because the calibrator itself exhibits circumstellar emission, or because of a decrease in atmospheric coherence time \citep{ertel_2016}. However, this would result in a negative bias, rather than a the positive which we find. Interestingly, the approximately ten undetected excesses that we estimate above would yield a detection rate that is compabible with that reported by \citet{absil_2013}. Also, since the estimate of the instrumental noise from the negative significance values is consistent with a normal distribution, it is most likely that the origin of the positive mean circumstellar emission is indeed astrophysical.

\subsection{Significance distribution of the JouFLU extension and follow-up observations }

\begin{figure}
  \begin{center}
    \includegraphics[scale=0.45]{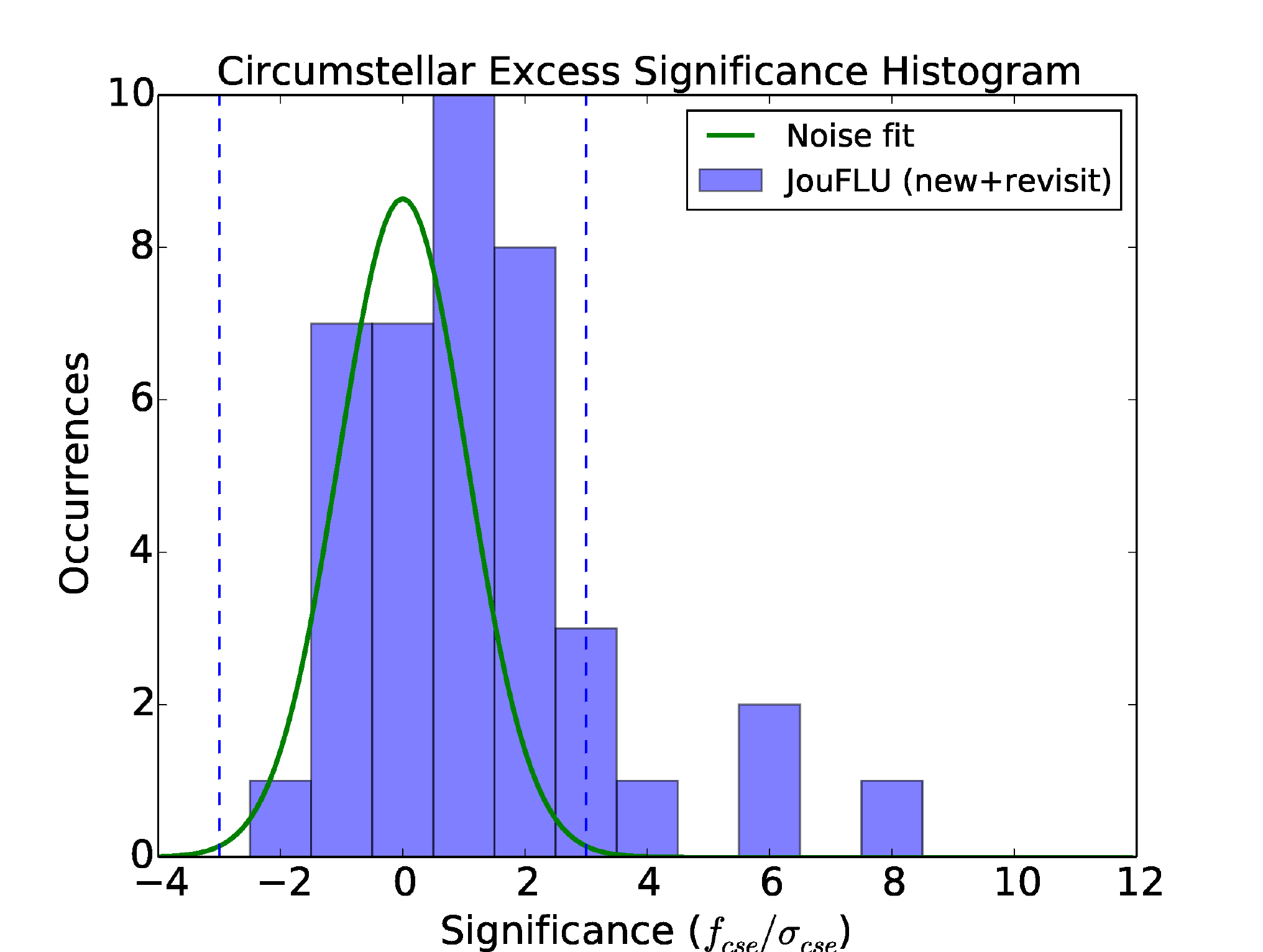}
  \end{center}
  \caption{\label{sig_histo_all_jouflu} Significance histogram that includes all the targets observed with JouFLU, including the follow-up targets of \citet{absil_2013} presented in Table \ref{follow_up_table}}
\end{figure}

Next, we examine all the targets observed with JouFLU, including the follow-up targets of the \citet{absil_2013} survey presented in Table \ref{follow_up_table}, yielding a total of 41 stars after removing all the excesses due to binarity. This is clearly a biased sample since most of the follow-up targets were previously reported as excess stars. For this biased sample, we report 7/41 excesses attributed to uniform circumstellar emission. The significance distribution of all of these JouFLU observations is shown in Fig. \ref{sig_histo_all_jouflu}. It has a median of 0.86, naturally more biased than the distribution of the JouFLU survey alone (Fig. \ref{sig_histo_old_new}). We again estimate the JouFLU instrumental noise by using the negative significance values (Fig. \ref{sig_histo_all_jouflu}), find the best fit standard deviation of the noise is 1.02. This is even closer to unity than with the JouFLU survey targets alone, again supporting our errorbar calculations.

\subsection{Significance distribution of calibrators} \label{cal_analysis}

\begin{figure*}
  \begin{center}
    \begin{eqnarray}
      \includegraphics[scale=0.45]{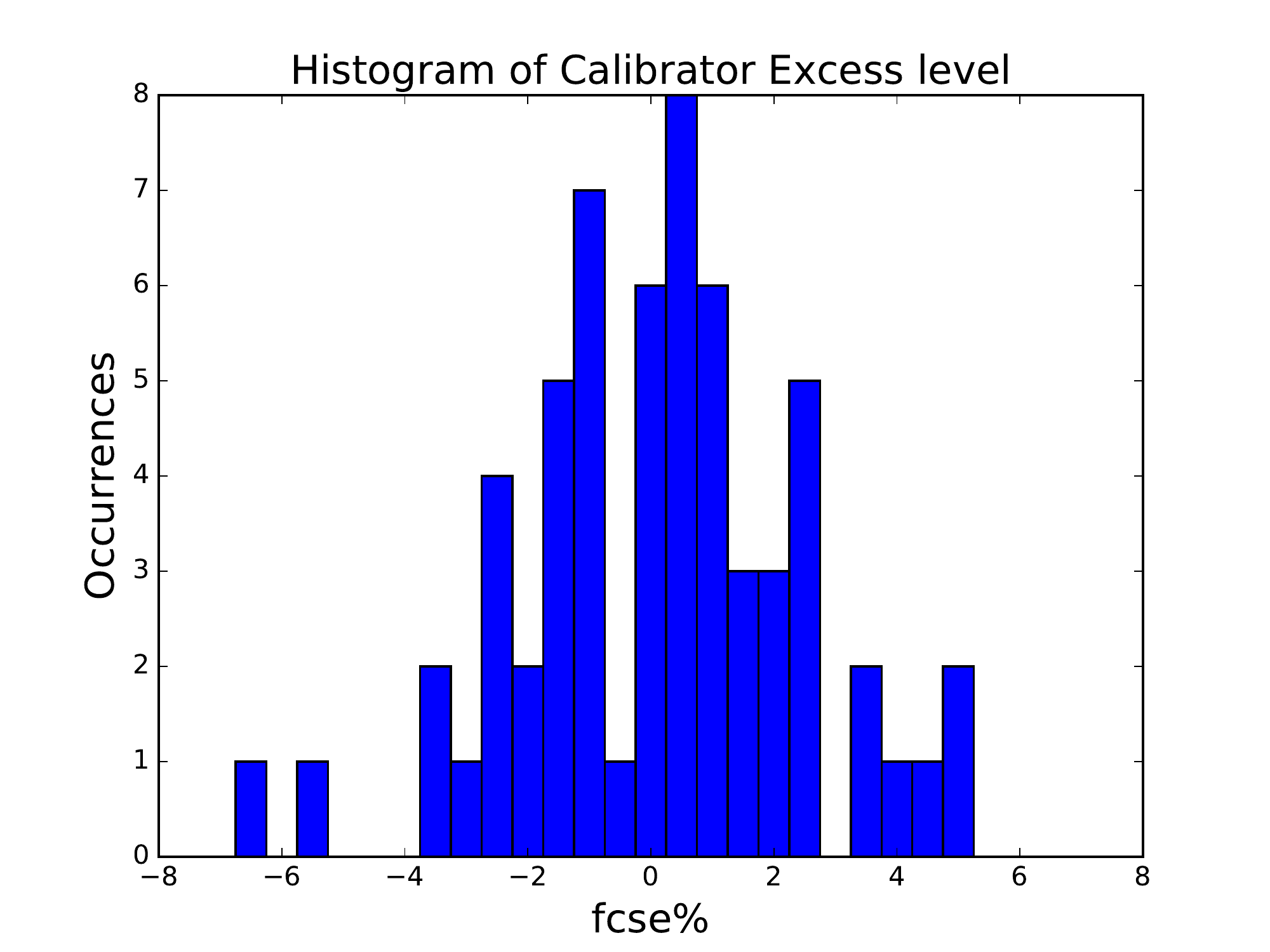}
      &  \includegraphics[scale=0.45]{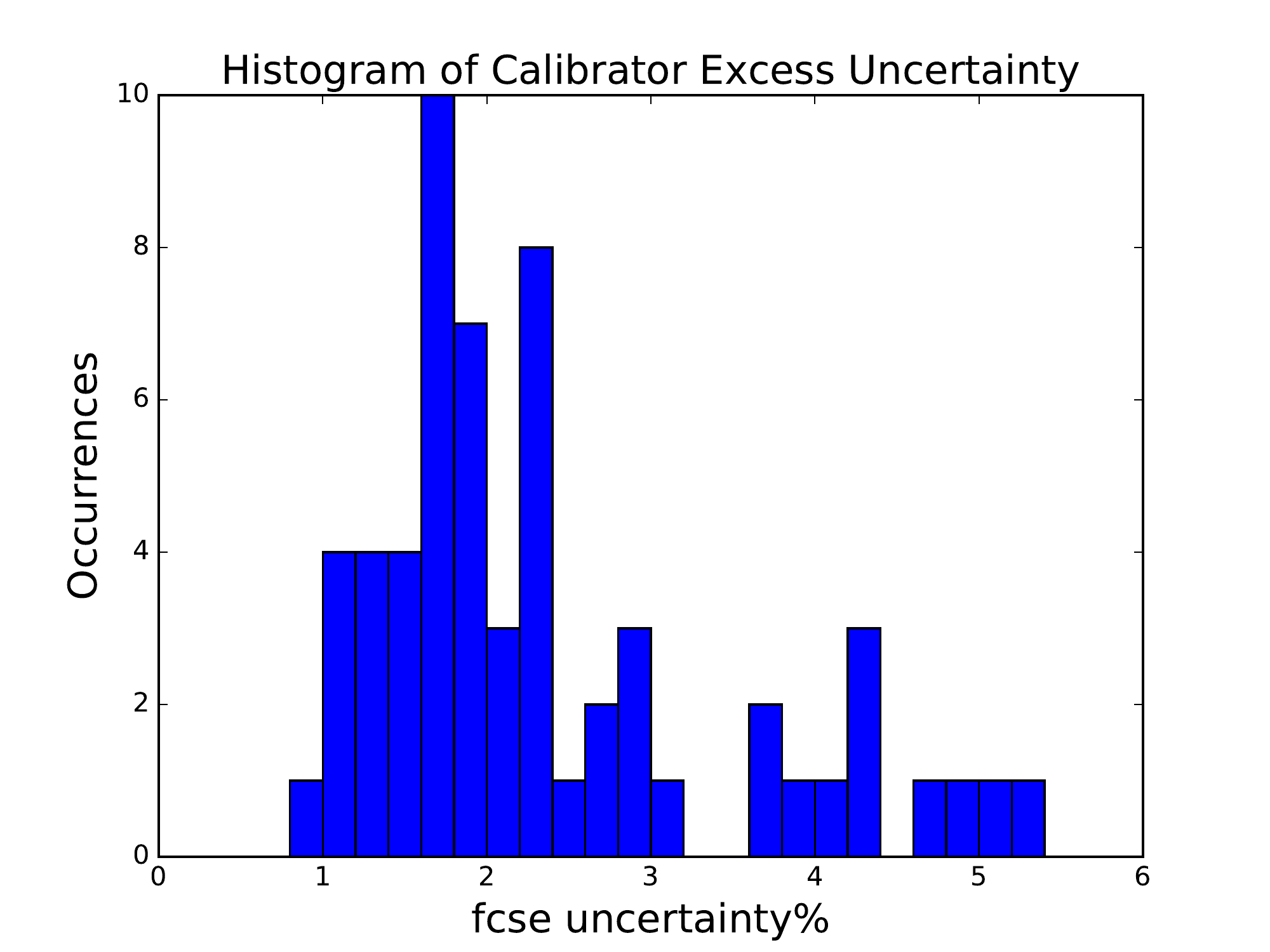} \nonumber
    \end{eqnarray}
  \end{center}
  \caption{\label{cal_fcse_histo} Circumstellar excess level of the calibrator stars used in this study (left panel) and its $1\sigma$ uncertainty (right panel). For each calibrator triplet, we used two calibrators to calibrate the remaining calibrator. We note that this distribution is unbiased, with a mean of $0.24\pm 0.37\%$.}
\end{figure*}

To further confirm the validity of our results, and in particular, to confirm that the positive mean of the significance distributions of the science targets is indeed of astrophysical origin, we computed the circumstellar excess levels of the calibrators. Since we nominally use three different calibrators for each science target, we can treat each calibrator as if it were a science target by calibrating it with the remaining two calibrators. So, for a set of three calibrators $cal_1$, $cal_2$, $cal_3$, we can compute the circumstellar excess of $cal_1$, by using $cal_2$ and $cal_3$ to calibrate $cal_1$, and similarly measure the excess of either $cal_2$ or $cal_3$.  In Fig. \ref{cal_fcse_histo} (left panel) we show the resulting circumstellar excess distribution of calibrators, and we note that this distribution is indeed unbiased, with a mean circumstellar excess of $(0.24\pm 0.37)\%$. For the particular case of the FLUOR follow-up targets, we find an unbiased mean circumstellar excess level of $(-0.21 \pm 0.73) \%$. From Fig. \ref{cal_fcse_histo} (right panel), which shows the uncertainty distribution of the circumstellar excess levels, we note that the median uncertainty is $2\%$, somewhat noisier than for the science targets because each calibrator was observed fewer times ($\sim 1/3$) than its corresponding science target.

\begin{figure}
  \begin{center}
    \includegraphics[scale=0.45]{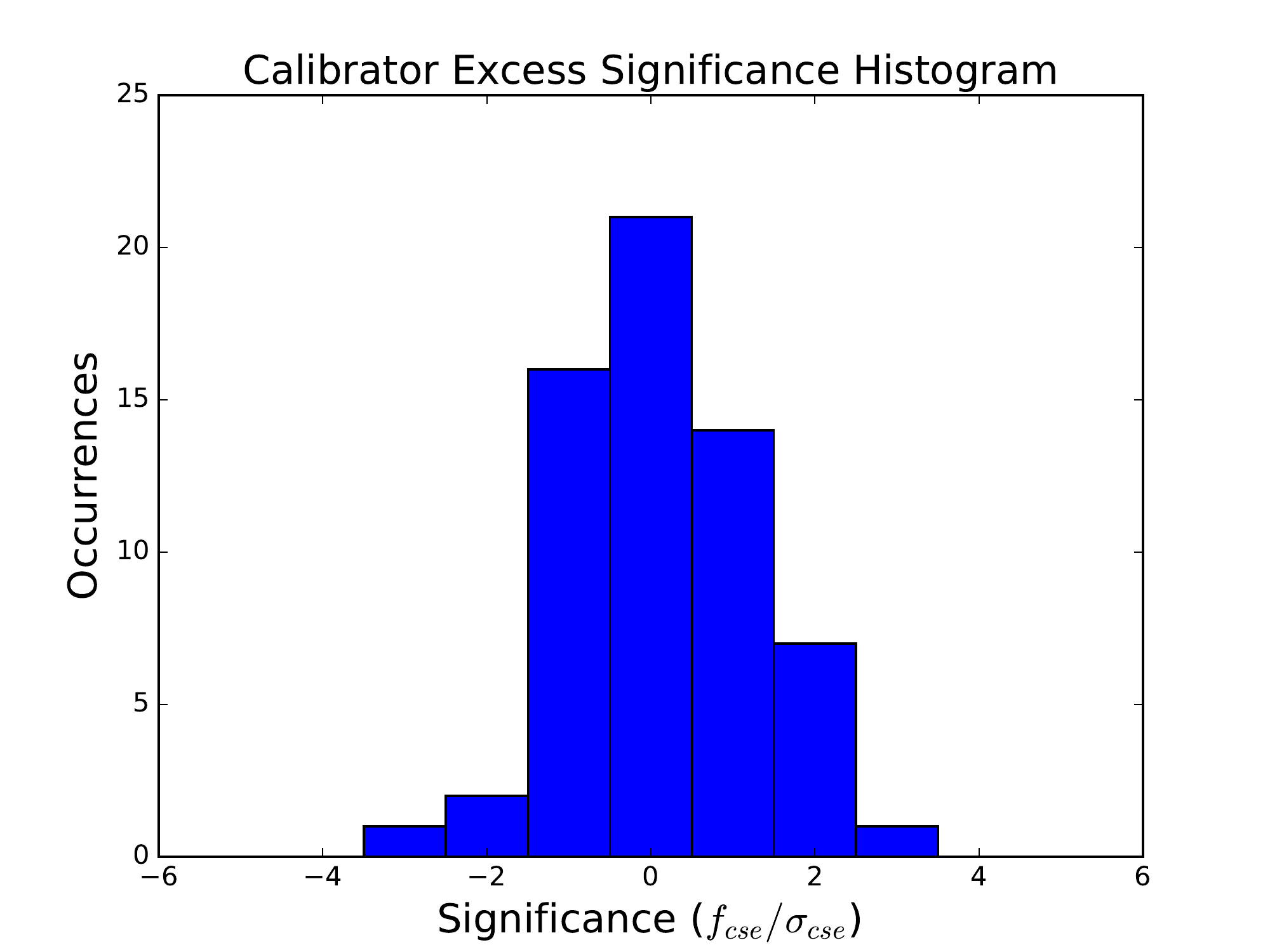}
  \end{center}
  \caption{\label{cal_sig_histo} Circumstellar excess significance level of the calibrator stars used in this study. We note that this distribution is unbiased, with a mean of 0.12, and a standard deviation of 1.03.}
\end{figure}

In Fig. \ref{cal_sig_histo} we show the significance distribution of the calibrator excess level, with a mean significance of $0.12$ and a standard deviation of $1.03$. We also find no calibrators with excesses above $3\sigma$ or below $-3\sigma$. In contrast, we do find several target stars above the $3\sigma$ detection threshold. The fact that we find no significant bias in this calibator analysis strongly supports the validity of our results.

\subsection{Significance distribution of the combined JouFLU and FLUOR samples} \label{combined}

\begin{figure}
\begin{center}
  \includegraphics[scale=0.45]{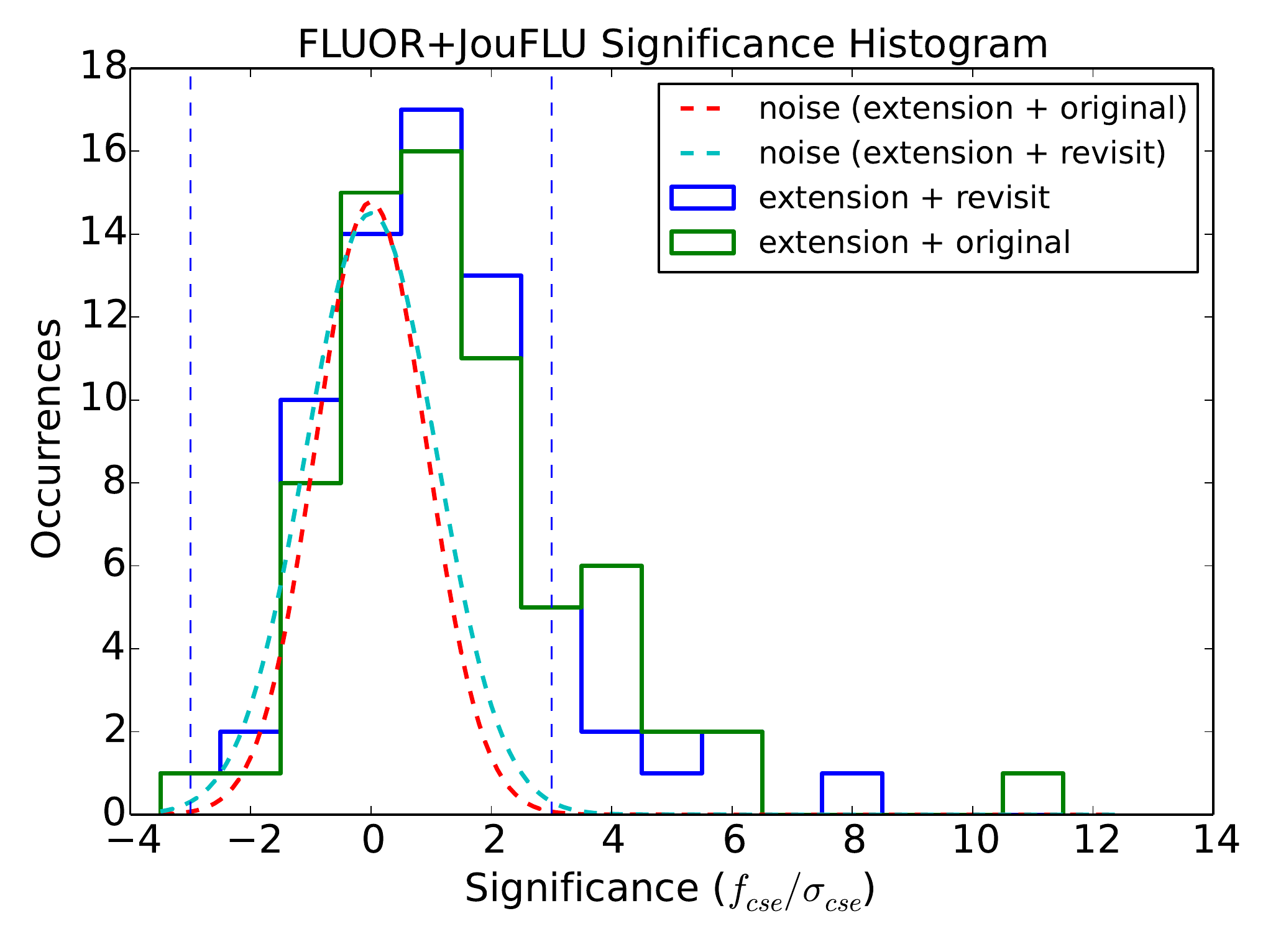}
\end{center}
\caption{\label{sig_fluor_jouflu} JouFLU and FLUOR samples combined. The blue histogram includes the JouFLU survey and the revisited (follow-up) targets with JouFLU. The blue histogram shows uses the original significance values for the follow-up targets obtained by \citet{absil_2013}. The dotted lines are the estimated instrumental noise significance distributions for the two different ways of combining the JouFLU and FLUOR observations. }
\end{figure}

Now we consider a couple of ways of combining the JouFLU and FLUOR samples, which differ in the way we include the follow-up targets. First,  we use the 40 FLUOR measurements and add the 29 single new star measurements from JouFLU. This is shown in Fig. \ref{sig_fluor_jouflu}, which has a corresponding detection rate of 15/69, or $21.7^{+5.7}_{-4.1}\%$. Second, we use the same sample of stars but now use the JouFLU measurements for the 11 FLUOR stars that JouFLU followed up, which results in a detection rate of 11/69, or $15.9^{+5.3}_{-3.8}\%$, as shown in Fig. \ref{sig_fluor_jouflu}. The detection rates for these two ways of combining the JouFLU and FLUOR samples are compatible with each other. That is, by computing the Kolmogorov-Smirnov statistic, we note that these two distributions are likely derived from the same distribution, since we obtain a p-value of 0.94. Also, according to the results of the Shapiro-Wilk test, the probabilities that these two distributions are derived from a normal distribution are $5.6\times 10^{-7}$ and $0.0120$ respectively. We also estimate the significance distributions of the noise, again by using the negative significance bins, and we find that the standard deviations using the old and new follow-up data are 1.07 and 0.92. Both are close to unity, as expected.

Since the FLUOR targets have detections with a higher signal-to-noise ratio, in the subsequent analyses of the joint FLUOR+JouFLU sample, we used the older FLUOR results along with results obtained for new targets by JouFLU, shown in Fig. 12.

\subsection{Correlations with spectral type and detected cold dust} \label{stype_corr}

\begin{figure*}
  \begin{center}
    \begin{eqnarray}
      \includegraphics[scale=0.4]{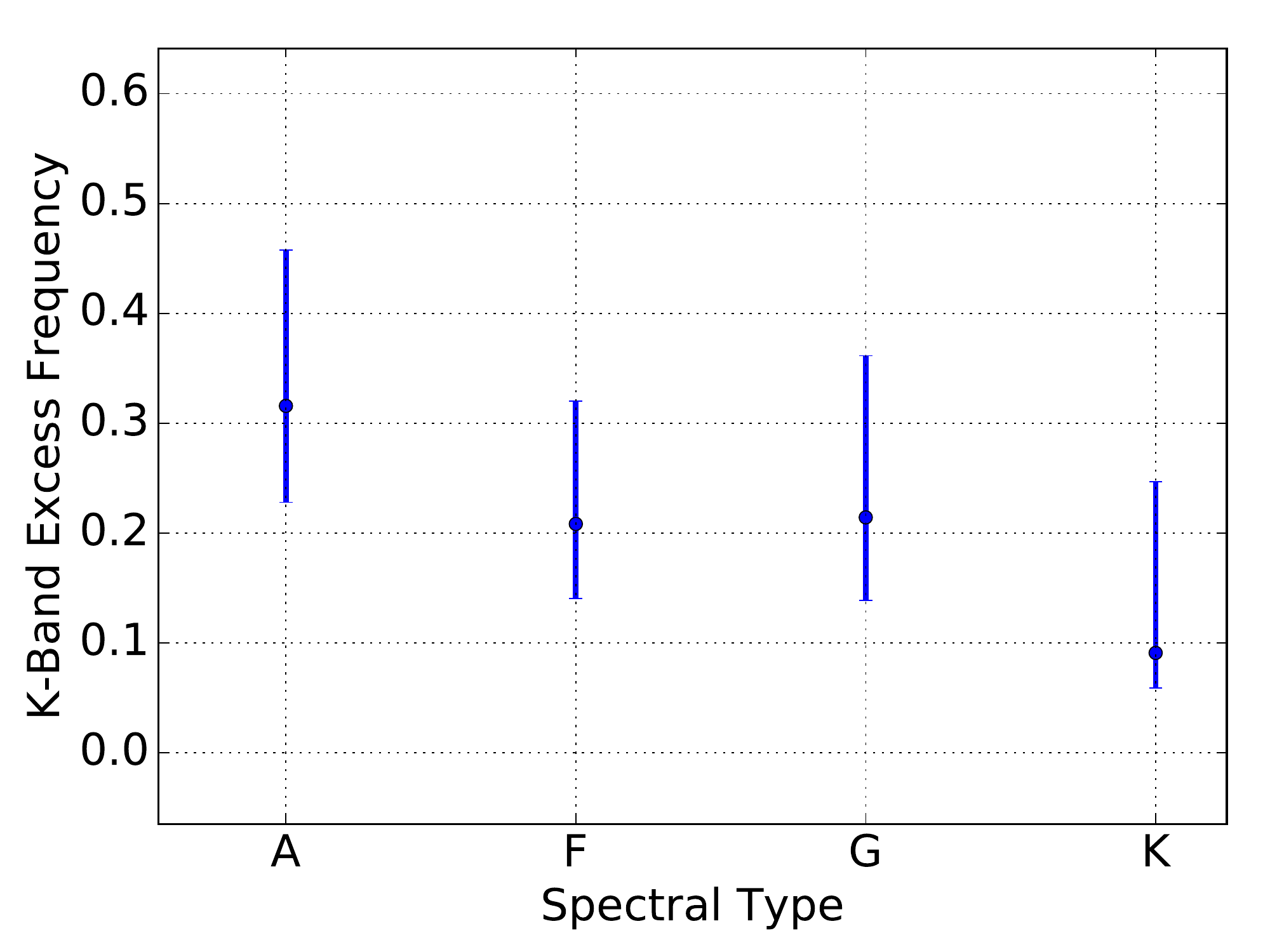} & \includegraphics[scale=0.42]{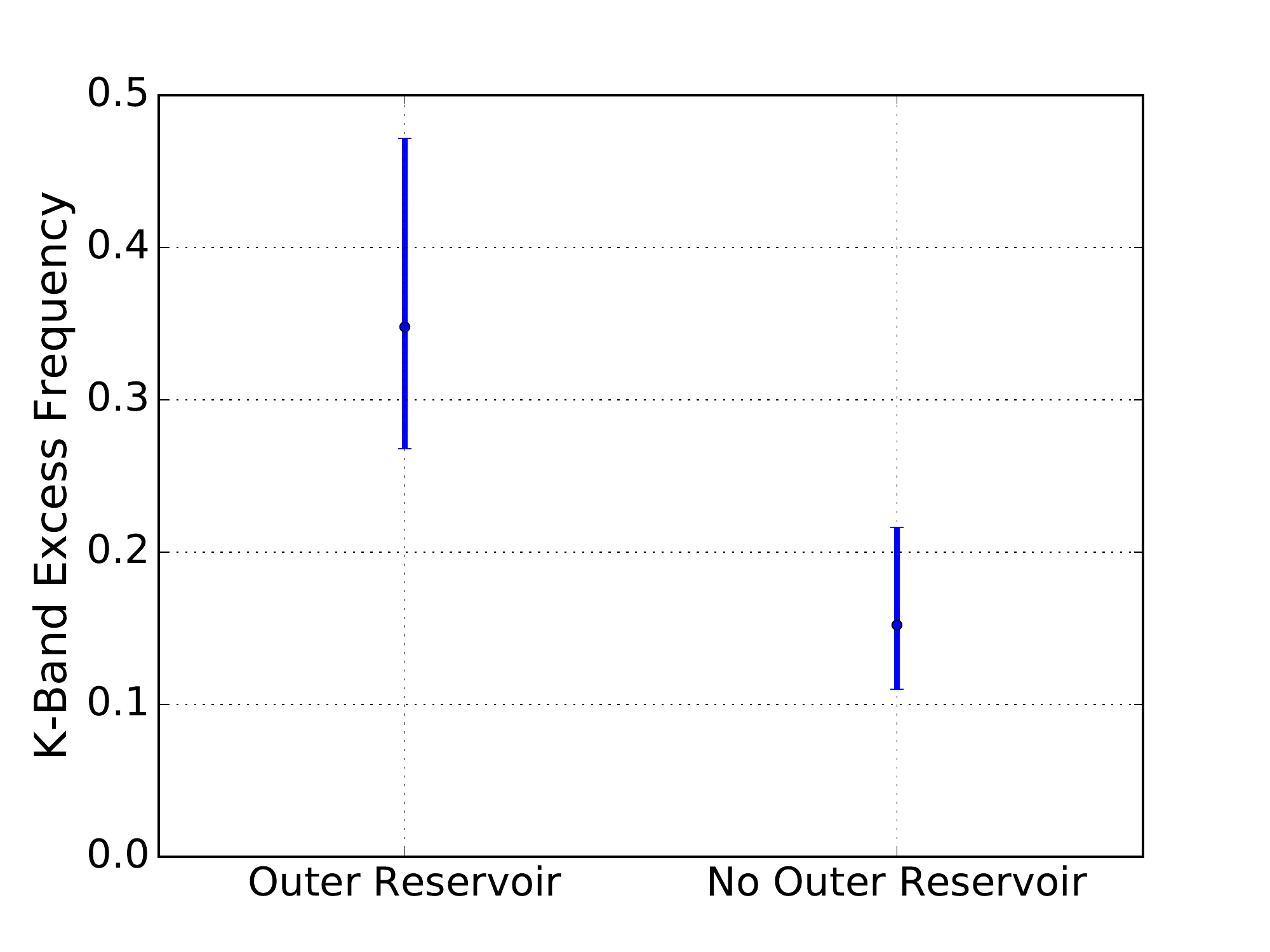} \nonumber
    \end{eqnarray}
  \end{center}
  \caption{\label{excess_stype_res}Excess rates of Table \ref{stype_res_det}. Left panel: excess rate for different spectral types. Right panel: excess rate with or without a detected cold-dust reservoir. The error-bars are asymmetric, are and computed by numerically integrating the binomial distribution. }
\end{figure*}

Now we look at how the detection rate of 15/69 is distributed for different spectral types and the presence of an outer cold-dust reservoir as shown in Table \ref{stype_res_det} and Fig. \ref{excess_stype_res}. In Fig. \ref{excess_stype_res}, we can see that the K-band excess occurrence rates for all spectral types are compatible with each other, showing a small decrease in excess rate for later-type stars, but not statistically significant. This tentative trend was also seen in the H-band interferometry data by \citet{ertel_2014}. It is reminiscent of the spectral dependency seen in far infrared excess rates, which are tracing the presence of colder outer debris disks  \citep{su_2006, bryden_2006}. The initial K-band FLUOR data had a significantly higher detection rate for A-type stars \citep{absil_2013},  but this is no longer the case when the FLUOR and JouFLU samples are combined. This could be an effect of the increased error bars of the JouFLU survey. 

\begin{table}
  \caption{\label{stype_res_det} K-band excess detection rate for stars of different spectral types, with and without an outer cold dust reservoir (as inferred from a MIR or FIR excess)}
  \begin{center}
    \begin{tabular}{ccccc}
      \hline\hline
                         &  A   &  F   &  G-K   &  Total \\\hline
      Outer res          & 2/9  &  4/9 &  2/5   &  8/23  \\
      No outer res       & 4/11 &  1/15 &  2/20 &  7/46 \\ \hline
      Total              & 6/20 &  5/24 &  4/25 &  15/69   \\ \hline
    \end{tabular}
  \end{center}
\end{table}

\begin{figure}
  \begin{center}  
    \includegraphics[scale=0.45]{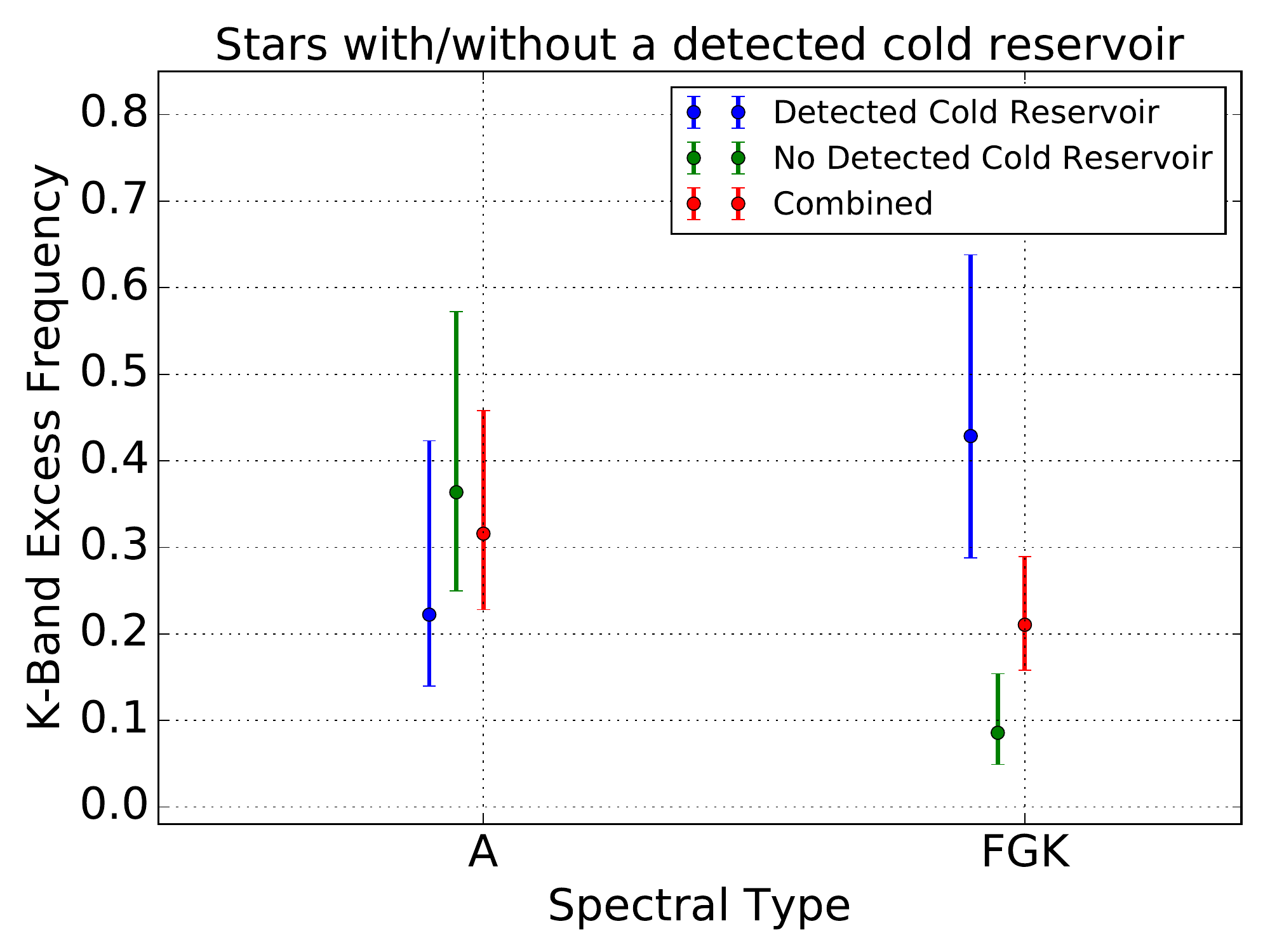}
  \end{center}
  \caption{\label{excess_stype_res1} Excess rates for different spectral types. The blue points represent excesses that have a corresponding detected cold reservoir. The green points represent excesses that do not have a detected cold reservoir. The red points represent the combined excesses with and without detected cold dust. Note that FGK stars have higher excess rate when they are known to have a bright  reservoir of cold dust.}
\end{figure}

From Fig. \ref{excess_stype_res} we also note that the excess rate for stars with a detected outer reservoir is higher, but once again, this is not statistically significant. However, if we split the cold-dust and no-cold-dust detections into their respective spectral types, as shown in Fig. \ref{excess_stype_res1}, we note that FGK-type stars tend to display excesses more frequently when they have a corresponding cold dust reservoir, while the same cannot be said about A-type stars. To quantify the significance of this trend, we assume binomial statistics for the excess rates shown in Table \ref{stype_res_det}, and we find that there is a 99\% probability that FGK stars display an excess more frequently when they have a known cold reservoir, than when they do not have a known cold reservoir. This trend was also present in the initial FLUOR survey performed by \citet{absil_2013}. The above analysis leads us to suspect that the mechanism responsible for the circumstellar excess is different for A-type stars than for FGK stars; or that there are two mechanisms behind A-type excesses, and one for FGK excesses.

\subsection{Correlations with rotational velocity}

\begin{figure}
  \begin{center}
    \includegraphics[scale=0.45]{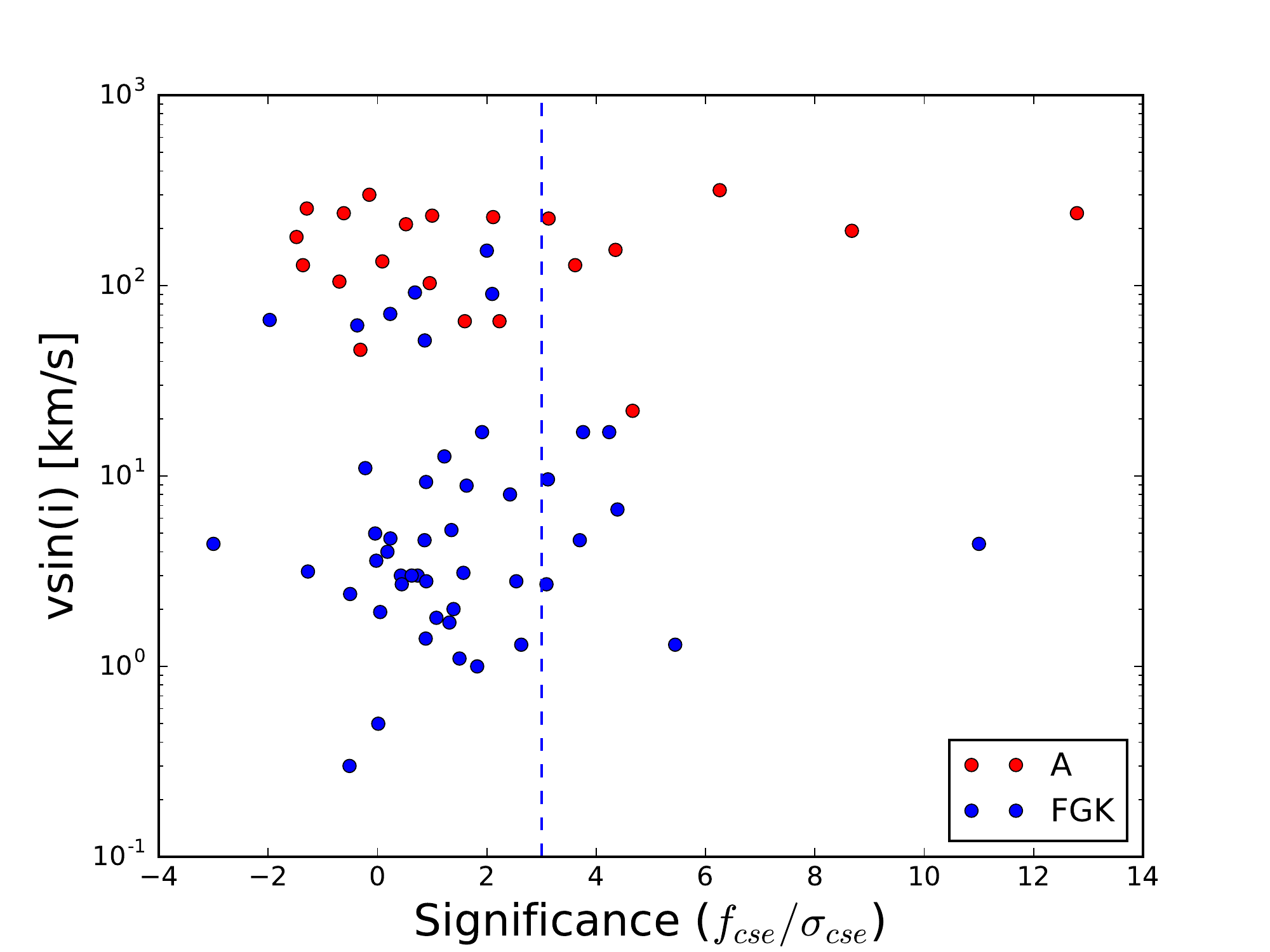}
  \end{center}
  \caption{\label{vsini_fig} Rotational velocity ($v sini$) as function of detection significance ($f_{\mathrm{cse}}/\sigma_{\mathrm{cse}}$). Red points correspond to A-type stars, and blue points correspond to FGK-type stars.}
\end{figure}

Next we investigated the excess rate as a function of the rotational velocity, or rather its apparent rotational velocity $v\sin i$, given the system's inclination angle $i$. From Fig. \ref{vsini_fig}, it is not obvious that detections favor either higher or lower values of $v\sin i$. However, we note a small accumulation of non-detections at lower values of $v\sin i$  as can be seen in the figure. If we take the median values of $v\sin i$ for the 14 detections and the 54 non-detections we find

\begin{eqnarray}
  \mathrm{Median}&(v\sin i)_{\mathrm{non-det}}= 6.6\pm^{4.4}_{2.1}\mathrm{km/s}\\ \nonumber
  \mathrm{Median}&  \!\!\!\!\!\!\!\!\!\!\!(v\sin i)_{\mathrm{det}}= 17 \pm^{10.3}_{5.0}\mathrm{km/s},
\end{eqnarray}

where the uncertainties for the median where estimated by generating many bootstrapped $v\sin i$ samples, and finding the $68\%$ confidence interval of the distribution of bootstrapped medians. The median value of $v\sin i$ for the detections is higher than for the non-detections, and we can estimate the significance of this result using bootstrapping: we generated many bootstrap samples of the detections and non-detections and count the number of times that the median $v\sin i$ of the detections is higher than for the non-detections. We find a probability of $\sim 88\%$ that the median $v\sin i $ for detections is larger than the median $v\sin i$ of non-detections. While this is not highly significant, it is an interesting trend that we can further explore by adding the 92 stars observed by \citet{ertel_2014}. In this case we find $\mathrm{median}(v\sin i)_{\mathrm{non-det}}= 9.1\pm^{2.4}_{1.9}\mathrm{km/s}$ and $\mathrm{median}(v\sin i)_{\mathrm{det}}= 16 \pm^{6.0}_{3.0}\mathrm{km/s}$, and the probability that excess stars have a higher median $v\sin i$ now increases to $91\%$. This statistical trend is particularly interesting in view of the recent theoretical study performed by \citet{rieke_2016}, which attributes the hot dust phenomenon to magnetic trapping of dust nano-grains close to the star, without the need for very high magnetic fields, and predict that there should be a correlation with high rotational velocities. We also note that there are a wide variety of stellar spectral types and hence rotational velocity ranges in the detection and non-detection samples. So to verify that this trend is not an effect of spectral type rather than rotational velocity, we compute the median spectral type of detections and non-detections, and find F6 for F6.5 respectively, which are very close to each other.

\subsection{Other possible correlations}

We have also investigated possible correlations of the excess levels with the existence of detected exoplanets, but have not found such a correlation. There are 8 known exoplanet host stars in the FLUOR and JouFLU combined samples: HD142091, HD219134, HD217014, HD190360, HD117176, HD69830, HD20630, and HD9826. We detect a K-band circumstellar excess for two such planet host stars: HD9826 ($\upsilon$ And) and HD142091 ($\kappa$ CrB). The other stars are consistent with a non-excess, and with good fits to stellar photosphere models. Last, we investigated possible correlations with stellar age. Similar to \citet{absil_2013} and \citet{ertel_2014}, we do not see any significant correlation with stellar age.

\section{Discussions and conclusions} \label{discussion}

The CHARA/JouFLU beam combiner was used to extend the initial CHARA/FLUOR survey and search for exozodiacal light originating from within a few AU from main sequence and sub-giant stars. In addition, JouFLU performed follow-up observations on most of the FLUOR excess stars to search for possible variability in the circumstellar emission level. 

By extending the survey, JouFLU observed targets that were fainter in the K band by a magnitude of approximately one. We report 6/33 new circumstellar excesses at the $\sim 1\%$ level or higher, relative to the star, out of which two detections are attributed to uniform circumstellar emission, and the other four being known or suspected binaries. Our detection rate of 2/29, after removing binaries from the sample, is clearly incompatible with the rate of 12/41 reported earlier by \citet{absil_2013}, but this is most likely an effect of our degraded precision by a factor of between approximately two and three. Indeed, we note that if we artificially increase the original FLUOR error bars by a factor of two, we obtain a detection rate of 3/41, which is compatible with the JouFLU detection rate presented here. The fact that we are still detecting K-band excesses for some of the targets originally observed by \citet{absil_2013}, with a modified instrument, and new data reduction software, is supporting evidence that these K-band excesses are real, and not an instrumental or data reduction artifact. Also, the fact that our data for binaries with known orbital parameters are compatible with expected values strengthens the reliability of our results. After combining the FLUOR and JouFLU samples as described in Section \ref{combined}, we have a detection rate of 15/69 or $21.7^{+5.7}_{-4.1}\%$.

Since there is some disagreement between the JouFLU and FLUOR results, it is worth discussing some potential causes:
\begin{enumerate}
\item Underestimation of the circumstellar excess uncertainty.\\
  As discussed in Section \ref{statistical_analysis}, we estimated the instrumental noise significance distribution and verified that it is indeed consistent with a normal distribution. An underestimation of the circumstellar excess uncertainty would result in a standard deviation greater than 1 in the instrumental noise significance distribution, which is not the case.
\item A systematic bias in the measured excess level. \\
  As discussed in Section \ref{cal_analysis}, we measured the circumstellar excess level of the calibrators, and showed that there is no statistically significant bias in their excess levels. For the follow-up targets, and in particular for the targets whose excess level changed between the JouFLU survey and the FLUOR survey, we verified that calibrators could be calibrated against each other.
\item Intrinsic variability in the near-infrared excess.\\
  Since we found no evidence for potential causes 1 and 2, we attribute the discrepancies between the JouFLU and FLUOR results to intrinsic variability which we further discuss below.
\end{enumerate}

Among the JouFLU circumstellar excess detections which seem to display variability are two exoplanet host stars: HD9826 ($\upsilon$ And) with a measured circumstellar excess of $(3.62\pm0.61)\%$, and HD142091 ($\kappa$ CrB) with an excess of $(3.4\pm 0.52)\%$. Both of these systems had been previously detected by \citet{absil_2013} with FLUOR, although with a much lower circumstellar emission level, namely $(0.53\pm0.17)\%$ for HD9826, and $(1.18\pm0.19)\%$ for HD142091. These systems have been observed extensively with other techniques, that allow excluding the possibility that a faint stellar companion could account for the measured JouFLU and FLUOR excesses. Therefore, these excesses are most likely due to extended emission, which we attribute to hot dust, and point to large dynamical activity in these systems. We note that the typical orbital period of dust at the sublimation radius is of the order of a few days, so there is really no reason to expect a constant excess measurement between the FLUOR and JouFLU observations, which were made a few years apart. These two stars should be considered as high priority targets  for follow-up observations using high contrast high resolution instruments in the near- to mid-infrared.  

The origin of these excesses remains difficult to explain. A possible explanation, alternative to hot dust, is that these excess stars have undetected faint companions. A-type stars are more likely to have undetected faint compaions, as they show fewer spectral lines than cooler stars and also tend to be more active, making spectroscopic searches for companions more difficult. An interferometric search for faint companions on the \citet{ertel_2014} sample was carried out by \citet{marion_2014} using precise closure phase measurements. There are 92 stars in the sample of \citet{ertel_2014}, among which 30 are A-type stars, and the five new binaries found in the survey were all discovered around A-type stars. This binary detection rate of 5/30 is actually compatible with the (FLUOR+JouFLU) detection rate of 6/20 for A-type stars (within $\sim 1\,\sigma$). Additionally, in the particular case of A stars, we found that there was no correlation with a detected cold-dust reservoir (Sec. \ref{stype_corr}). So it is possible that the excesses detected around A-type stars are simply due to binarity. However, even if we assume that all of the JouFLU+FLUOR A-type excesses (6/20) are due to binarity, which is very unlikely for the well studied ones such as Vega and Altair, we are still left with 8/48 detections for all the other spectral types, still a significant detection rate. 

\begin{figure}
  \begin{center}
    \includegraphics[scale=0.45]{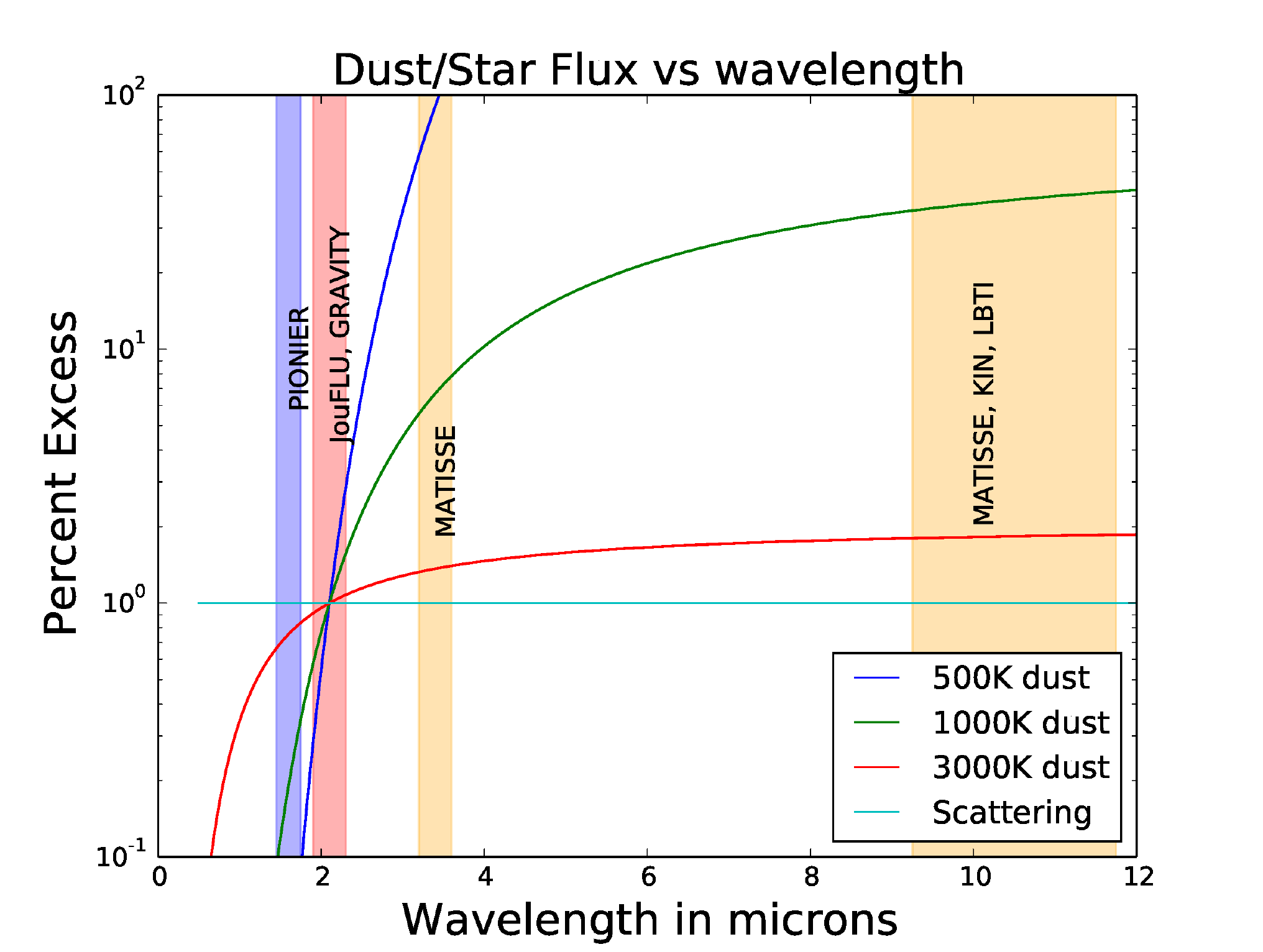}
  \end{center}
  \caption{\label{dust_curves} Assuming thermally emitting dust, with a K-band flux ratio of $1\%$ relative to a star of $6000\,$K, the curves show the expected contrast as a function of wavelength, and each curve corresponds to a different dust temperature. The shaded regions correspond to the accessible spectral bands for different interferometers, including second-generation instruments such as GRAVITY \citep{gravity_ref} and MATISSE \citep{matisse_ref}. }
\end{figure}

If we assume that the circumstellar excesses are not due to binarity, then they are most likely due to extended emission, which we attribute to hot dust. Assuming a dust sublimation temperature of $\sim 1500\,\mathrm{K}$, we estimate the dust sublimation radius to be between $\sim 2-30\,\mathrm{mas}$ for the JouFLU+FLUOR stellar sample, so the CHARA baselines should allow resolving dust for all targets. The dust properties have been somewhat constrained by other observations at different wavelengths and spatial resolutions. In Figure \ref{dust_curves} we show the expected K-band excess for thermally emitting dust with a K-band flux ratio of $1\%$ in the K band, relative to a $6000\,$K star, and assuming that dust is not confined to be at a certain distance from the star. According to the Figure, dust with gray blackbody emission in the near to MIR, should be more easily detected at longer wavelengths, for example $\sim 10\mu\mathrm{m}$, the central wavelength of the Keck Interferomer Nuller (KIN). \citet{mennesson_2014} used the KIN to search for exozodiacal light in the $8\mu\mathrm{m}-10 \mu\mathrm{m}$ band, and their sample included the FLUOR K-band excess stars detected by \citet{absil_2013}. Even though the KIN excess levels predicted at 10$\,\mu$m for gray emission dust are higher than in the near-infrared, none of the FLUOR excess stars showed a significant excess at 10$\,\mu$m. The spatial resolution of the KIN ($\sim$10$\,$mas) should also have allowed to resolve such dust in most cases. This implies a very high dust temperature ($>1000\,$K), with grains located close to the dust sublimation radius ($\sim 0.1\,$AU), and with typical grain sizes smaller than the blowout-size ($\sim 1\mu\mathrm{m}$) to escape detection in the mid-infrared; \emph{or} the dust emission is due to scattering, in which case it is more difficult to constrain its distance to the star. However, scattering seems less likely in view of the lower excess detection rate in the H band of $10.6^{+4.3}_{-2.5}\%$ \citep{ertel_2014}, and in view of polarimetric observations performed by \citet{marshall_2016}, which found no evidence of scattered light emitted from the NIR excess stars. \citet{kirchschlager_2017} modeled many hot dust systems, including KIN upper limits, and found that a population of reasonably hot, small grains is well consistent with the KIN results. They also found that a fraction of the light can indeed be scattered light for their realistic grains which would still be consistent with a low polarization fraction from \citet{marshall_2016}. In the case of Vega, the dust location has been constrained using the Palomar Fiber Nuller, a near-infrared interferometer with a $3.4\,$m baseline, to be either within $0.2\,$AU, or beyond $2\,$AU \citep{mennesson_2011}.

Close-in sub-micron dust should have short lifetimes of the order of years \citep{wyatt_2008}, so if dust is really responsible for the K-band excesses, then there must be an efficient replenishing or dust trapping mechanism. Several models exist to explain the persistence of dust, the most prominent are: comet in-fall and out-gassing \citep{wyatt_2008, su_2013, lebreton_2013}, or magnetic trapping of small grains \citep{rieke_2016}. The JouFLU+FLUOR data allowed to test for the magnetic trapping model of \citet{rieke_2016}, which predicts a correlation with high rotational velocities, and the data show a tentative correlation with angular velocity, but more data are needed to confirm this statistical trend. In order to test for episodic events, we are currently performing an exozodi monitoring program with JouFLU/CHARA, led by Nicholas Scott, where we are revisiting excess targets every few months to search for variability in the circumstellar emission level. 

It has been known for many years that a density enhancement can occur near the dust sublimation radius; as particles sublimate, the effect of stellar radiation pressure grows, slowing and eventually reversing their migration \citep{mukai_1979}. The effect of sublimation on the particle orbits has been explored in detail with numerical and analytic models \citep{kobayashi_2008, kobayashi_2009, kobayashi_2011}. The optical depth of the disk increases by factors of $\sim 3-10$ near the sublimation radius, depending on the grain composition and stellar spectral type. This increase in dust is promising, but is still not enough to explain the observations.

A possibility that is currently being investigated is the effect of the gas that is produced by the sublimation. Gas can further exaggerate the inner accumulation of dust (and thereby help explain the observations) in several ways. Firstly, the gas will circularize the eccentric orbits produced as the particles sublimate. This circularization acts to increase the outward migration already found in the Kobayashi models, helping the dust to pile up somewhat farther outward and to sublimate more slowly. Second, there will be direct gas drag on the dust particles. While gas drag is often assumed to be inward (e.g., for dust in thick protostellar disks), in the gas and dust environment considered here, drag forces actually push the dust outward (with radiation pressure offsetting the central gravity, small particles orbit slower, not faster, than the gas). Lastly, the gas acts to regularize the dust orbits, thereby lowering both their collision rate and their relative velocities. As such, the combined effect of these gas forces serves to not only enhance the amount of hot dust, but also to decrease its removal rate, allowing for a smaller supply of mass to produce strong emission over the lifetime of a star.

\section*{Acknowledgments:}
This research was supported by an appointment to the NASA Postdoctoral Program at the Jet Propulsion Laboratory administered by Universities Space Research Association under contract with NASA. PN and BM are grateful for support from the NASA Exoplanet Research Program element, though grant number NNN13D460T.
This work is based upon observations obtained with the Georgia State University Center for High Angular Resolution Astronomy Array at Mount Wilson Observatory. The CHARA Array is supported by the National Science Foundation under Grant No. AST-1211929. Institutional support has been provided from the GSU College of Arts and Sciences and the GSU Office of the Vice President for Research and Economic Development.
This research has made use of the Simbad database, operated at the Centre de Donn\'ees Astronomiques de Strasbourg (CDS), and of NASA's Astrophysics Data System Bibliographic Services (ADS). Thanks to those developping the Aspro2 software and those maintaining the Jean Marie Mariotti Center (JMMC). We would also like to thank Elliot Horch for providing a reduction of complementary speckle imaging data for HD9826 obtained from NESSI (NN-EXPLORE Exoplanet \& Stellar Speckle Imager). 

\bibliography{survey_bibliography}
\end{document}